\documentclass[a4paper,12pt]{article}
\usepackage{amssymb}
\usepackage{amsmath}
\usepackage{amsfonts}
\usepackage{mathtools}
\usepackage{bm}
\usepackage{hyperref}
\usepackage{slashed}
\usepackage{color}
\usepackage{indentfirst}
\usepackage{graphicx}
\usepackage{bbm}
\usepackage{csquotes} 
\usepackage{caption}
\usepackage{here}
\usepackage{comment}
\usepackage{listings}
\usepackage[italicdiff]{physics}
\numberwithin{equation}{section} 

\usepackage[
backend=biber,
eprint=true,
sorting=none,
sortcites=true,
maxnames=8,
date=year
]{biblatex}
\def\ignore#1{{}}

\makeatletter
\newcommand*\rel@kern[1]{\kern#1\dimexpr\macc@kerna}
\newcommand*\widebar[1]{%
  \begingroup
  \def\mathaccent##1##2{%
    \rel@kern{0.8}%
    \overline{\rel@kern{-0.8}\macc@nucleus\rel@kern{0.2}}%
    \rel@kern{-0.2}%
  }%
  \macc@depth\@ne
  \let\math@bgroup\@empty \let\math@egroup\macc@set@skewchar
  \mathsurround\z@ \frozen@everymath{\mathgroup\macc@group\relax}%
  \macc@set@skewchar\relax
  \let\mathaccentV\macc@nested@a
  \macc@nested@a\relax111{#1}%
  \endgroup
}
\makeatother

\renewcommand{\thefootnote}{\arabic{footnote}}
\clearpage
\addtocounter{page}{-1}

\def\ignore#1{{}}

\newcommand{\alp}{\alpha}

\newcommand{\Gm}{\Gamma}
\newcommand{\dlt}{\delta}
\newcommand{\Dlt}{\Delta}
\newcommand{\ep}{\epsilon}
\newcommand{\tht}{\theta}

\newcommand{\vth}{\vartheta}
\newcommand{\kp}{\kappa}
\newcommand{\lmd}{\lambda}
\newcommand{\Lmd}{\Lambda}
\newcommand{\sgm}{\sigma}

\newcommand{\omg}{\omega}

\newcommand{\bea}{\begin{eqnarray}}
\newcommand{\eea}{\end{eqnarray}}

\newcommand{\simgt}{\stackrel{>}{{}_\sim}}

\newcommand{\tl}[1]{\tilde{#1}}

\newcommand{\bdm}[1]{{\bm #1}}

\newcommand{\diag}{{\rm diag}}
\newcommand{\der}{\partial}

\newcommand{\id}{\mbox{\boldmath $1$}}

\newcommand{\asc}{a_{\rm sc}}
\newcommand{\pf}{p_{\rm 4d}}
\newcommand{\pth}{p_{\rm 3d}}
\newcommand{\tpt}{\tilde{p}_{\rm 3d}}
\newcommand{\Lct}{\Lambda_{\rm cut}}
\newcommand{\yh}{y_{\rm h}}

\newcommand{\Lvev}[1]{\left\langle #1 \right\rangle}
\newcommand{\brkt}[1]{\left( #1 \right)}
\newcommand{\brc}[1]{\left\{ #1 \right\}}
\newcommand{\sbk}[1]{\left[ #1 \right]}

\newcommand{\udl}[1]{\underline{#1}}

\newcommand{\cC}{{\cal C}}
\newcommand{\cD}{{\cal D}}
\newcommand{\cE}{{\cal E}}
\newcommand{\cF}{{\cal F}}
\newcommand{\cG}{{\cal G}}
\newcommand{\cH}{{\cal H}}
\newcommand{\cI}{{\cal I}}

\newcommand{\cK}{{\cal K}}
\newcommand{\cL}{{\cal L}}

\newcommand{\cO}{{\cal O}}

\newcommand{\cR}{{\cal R}}
\newcommand{\cS}{{\cal S}}
\newcommand{\cT}{{\cal T}}
\newcommand{\cU}{{\cal U}}
\newcommand{\cV}{{\cal V}}
\newcommand{\cW}{{\cal W}}

\renewcommand{\thefootnote}{\fnsymbol{footnote}}

\begin{document}

\title{
\begin{flushright}
\begin{minipage}{0.25\linewidth}
\normalsize
KEK-TH-2773 \\
\end{minipage}
\end{flushright}
{\Large \bf 

Quantum vacuum energy \\
and geometry of extra dimension
\\*[20pt]}}

\author{
Yutaka~Sakamura\footnote{
E-mail address: sakamura@post.kek.jp
}\\*[20pt]
{\it \normalsize 
KEK Theory Center, Institute of Particle and Nuclear Studies, KEK,}\\
{\it \normalsize 1-1 Oho, Tsukuba, Ibaraki 305-0801, Japan}\\
{\it \normalsize 
Graduate University for Advanced Studies (Sokendai),}\\
{\it \normalsize 1-1 Oho, Tsukuba, Ibaraki 305-0801, Japan.}
}
\date{}

\maketitle

\centerline{\small \bf Abstract}
\begin{minipage}{0.9\linewidth}
\medskip 
\medskip 
\small 
We discuss the cancellation of the ultraviolet cutoff scale~$\Lct$ in the calculation of the expectation value 
of the five-dimensional (5D) energy-momentum tensor~$\Lvev{T_{MN}}$ ($M,N=0,1,\cdots,4$). 
Since 5D fields feel the background geometry differently depending on their spins, the bosonic and 
the fermionic contributions to the $\Lct$-dependent part~$\Lvev{T_{MN}}^{\rm UV}$ may have different profiles 
in the extra dimension. 
In that case, there is no chance for them to be cancelled with each other. 
We consider arbitrary numbers of scalar and spinor fields with arbitrary bulk masses, 
calculate $\Lvev{T_{MN}}$ using the 5D propagators, 
and clarify the dependence of $\Lvev{T_{MN}}^{\rm UV}$ on the extra-dimensional coordinate~$y$ 
for a general background geometry of the extra dimension. 
We find that if the geometry is not flat nor (a slice of) anti-de Sitter (AdS) space, 
it is impossible to cancel $\Lvev{T_{MN}}^{\rm UV}$ between the bosonic and the fermionic contributions. 
This may suggest that the flat (or AdS) space is energetically favored over the other geometries, 
and thus the dynamics forces the compact space to be flat (or AdS). 
\end{minipage}

\renewcommand{\thefootnote}{\arabic{footnote}}
\thispagestyle{empty}
\clearpage

\section{Introduction}
\label{introduction}
The extra dimensions are predicted by string theory, and have been extensively investigated 
in many papers including their phenomenological impacts. 
If the compact extra space exists, there is a quantum contribution to the vacuum energy, which is similar to the Casimir energy~\cite{Casimir:1948dh}. 
Since it depends on the size of the compact space, it is often discussed in the context 
of the moduli stabilization~\cite{Garriga:2000jb,Toms:2000bh,Goldberger:2000dv,Brevik:2000vt}. 
There is also a possibility that this ``Casimir energy'' is the origin of the dark energy, 
which is recently discussed in the context of the dark dimension scenario~\cite{Montero:2022prj}.  

In general, the quantum correction to vacuum energy density~$\rho_{\rm eng}$ is divergent. 
In other words, it depends on the ultraviolet (UV) cutoff scale~$\Lct$. 
Thus it can be expressed in the form of $\rho_{\rm eng}=\rho_{\rm cas}(R)+\rho_{\rm div}(\Lct)$, 
where $R$ denotes the size modulus of the compact space, 
$\rho_{\rm cas}(R)$ is the modulus dependent part, which is called the Casimir energy, 
and $\rho_{\rm div}(\Lct)$ is the modulus independent part and of $\cO(\Lct^p)$ ($p\geq 1$). 
The most efficient way to calculate $\rho_{\rm cas}(R)$ is to use the regularization based on the analytic continuation, 
such as the dimensional regularization, the zeta-function regularization and so on~\cite{Beneventano:1995fh,Moretti:1998rf,Hagen:2000bu,Visser:2016ddm,Asai:2021csl}. 
In these regularization schemes, the $\Lct$-dependent part~$\rho_{\rm div}$ automatically dropped, and we can obtain 
the finite result. 
When we discuss the modulus stabilization, this derivation causes no problem because $\rho_{\rm div}$ is $R$-independent 
and irrelevant to the discussion. 
However, once the gravity is taken into account, the whole contribution~$\rho_{\rm eng}$ becomes relevant 
and $\rho_{\rm div}(\Lct)$ should not be neglected because what appears in the Einstein field equation is not $\rho_{\rm cas}$ but $\rho_{\rm eng}$. 
Usually, we implicitly assume some unknown mechanism that cancels $\rho_{\rm div}$ and the resultant vacuum energy density becomes 
$\rho_{\rm cas}$. 
One of the most plausible possibility is that 
the cancellation is realized among contributions of bosonic and fermionic particles, 
including undiscovered ones~\cite{Visser:2016mtr}. 
In this paper, we pursue this possibility. 

In the case that the size of the extra dimensions is large, such as the dark dimension scenario, 
we should work in the higher dimensional setup because 
the Kaluza-Klein (KK) mass scale can be lower than the temperature of the universe. 
In particular, when we discuss the evolution of the higher-dimensional bulk spacetime in the semiclassical approximation, 
we have to know the vacuum expectation value (VEV) of the higher-dimensional energy-momentum tensor~$\Lvev{T_{MN}}$. 
Similarly to the case of $\rho_{\rm eng}$ in the previous paragraph, 
$\Lvev{T_{MN}}$ can also be divided as $\Lvev{T_{MN}}=\Lvev{T_{MN}}^{\rm fin}(R)+\Lvev{T_{MN}}^{\rm UV}(\Lct)$. 
The four-dimensional (4D) energy density is related as $\rho_{\rm eng}=\int d^ny\,\Lvev{T_{tt}}$ 
and $\rho_{\rm cas}(R)=\int d^ny\,\Lvev{T_{tt}}^{\rm fin}(R)$, 
where $y^m$ ($m=1,2,\cdots,n$) are the coordinates of the extra dimensions. 
The purpose of this paper is to clarify how the cancellation of $\Lvev{T_{MN}}^{\rm UV}$ occurs. 
To simplify the discussions, we consider a five-dimensional (5D) theory compactified on an orbifold~$S^1/Z_2$. 
There is one concern about the cancellation. 
In contrast to $\rho_{\rm eng}$, the 5D quantity~$\Lvev{T_{MN}}$ can have a nontrivial dependence on the extra-dimensional coordinate~$y$. 
It is well known that the mode functions with a specific KK level have different profiles for 5D scalar, spinor and vector fields 
if the geometry is not flat~\cite{Gherghetta:2000qt}. 
In other words, 5D fields feel the background geometry differently depending on their spins. 
Therefore, a bosonic and a fermionic contributions to $\Lvev{T_{MN}}^{\rm UV}$ might have different profiles for $y$. 
If this is the case, it is impossible to cancel them with each other. 
We will clarify the contributions to $\Lvev{T_{MN}}$ from 5D scalar and spinor fields for a general background geometry 
of the extra dimension, and study the possibility of the $\Lvev{T_{MN}}^{\rm UV}$ cancellation. 

The paper is organized as follows. 
In the next section, we provide a brief review of the calculation of the vacuum energy density and the pressure in 4D effective theory, 
and derive the conditions required to cancel the $\Lct$-dependent part. 
In Sec.~\ref{5D_EMT}, we calculate the VEV of the 5D energy-momentum tensor~$\Lvev{T_{MN}}$ by using the 5D propagators. 
In Sec.~\ref{cutoff_dep}, we discuss the dependence of $\Lvev{T_{MN}}$ on the cutoff energy scale~$\Lct$, and its cancellation 
among bosons and fermions. 
Sec.~\ref{summary} is devoted to the summary and discussions. 
We have collected our notations in App.~\ref{notations}, 
the explicit component expressions of the Riemann tensor and the Einstein equation for our metric ansatz in App.~\ref{Einstein_eqs}, 
the definitions of the basis functions for solutions to the mode equations in App.~\ref{basis_fcns}, 
and the derivation of the 5D propagators in App.~\ref{derivation:5D_propagator}. 
In App.~\ref{LM_expansion}, we derive approximate expressions of various quantities for large momentum. 

\section{Vacuum energy density in 4D effective theory}
\label{review}
We begin with a brief review of the calculation of the vacuum energy density and the pressure in 4D effective theory. 
As a simple example, we consider a real scalar theory in a flat 5D spacetime compactified on $S^1/Z_2$. 
\begin{align}
 \cL_{\rm scalar} &= -\frac{1}{2}\der^M\Phi\der_M\Phi-\frac{M^2}{2}\Phi^2, 
 \label{cL:scalar}
\end{align}
where $M=0,1,2,3,4$. 
The fundamental region of $S^1/Z_2$ is chosen as $0\leq y\equiv x^4\leq \pi R$, where $R$ is the radius of $S^1$. 
The real scalar field~$\Phi$ is assumed to be $Z_2$-even. 
Then the vacuum energy density~$\rho_{\rm eng}$ and the vacuum pressure~$p_{\rm prs}$ are formally expressed as
\begin{align}
 \rho_{\rm eng} &= \sum_{n=0}^\infty \int\frac{d^3p}{2(2\pi)^3}\;\sqrt{p^2+m_n^2}
 = \sum_{n=0}^\infty \int_0^\infty \frac{dp}{4\pi^2}\;p^2\sqrt{p^2+m_n^2}, \nonumber\\
 p_{\rm prs} &= \frac{1}{3}\sum_{n=0}^\infty\int\frac{d^3p}{2(2\pi)^3}\;\frac{p^2}{\sqrt{p^2+m_n^2}} 
 = \sum_{n=0}^\infty\int_0^\infty \frac{dp}{12\pi^2}\;\frac{p^4}{\sqrt{p^2+m_n^2}}, 
 \label{expr:rho_eng}
\end{align}
where $m_n$ ($n=0,1,2,\cdots$) are the KK masses given by
\begin{align}
 m_n &= \sqrt{M^2+\frac{n^2}{R^2}}. 
\end{align}
Since this clearly diverges, it has to be regularized. 
Here we introduce the cutoff energy scale~$\Lct$, and the contributions above it are discarded. 
Then, \eqref{expr:rho_eng} becomes
\begin{align}
 \rho_{\rm eng} &= \sum_{n=0}^{N_{\rm cut}}\int_0^{\Lct}\frac{dp}{4\pi^2}\;p^2\sqrt{p^2+m_n^2} \nonumber\\
 &= \sum_{n=0}^{N_{\rm cut}}\sbk{\frac{\Lct^4}{16\pi^2}+\frac{\Lct^2m_n^2}{16\pi^2}+\frac{m_n^4}{128\pi^2}
 -\frac{m_n^4}{32\pi^2}\ln\frac{2\Lct}{m_n}+\cO\brkt{\frac{1}{\Lct^2}}}, \nonumber\\
 p_{\rm prs} &= \sum_{n=0}^{N_{\rm cut}}\int_0^{\Lct}\frac{dp}{12\pi^2}\;\frac{p^4}{\sqrt{p^2+m_n^2}} \nonumber\\
 &= \sum_{n=0}^{N_{\rm cut}}\sbk{\frac{\Lct^4}{48\pi^2}-\frac{\Lct^2m_n^2}{48\pi^2}
 -\frac{7m_n^4}{384}+\frac{m_n^4}{32\pi^2}\ln\frac{2\Lct}{m_n}+\cO\brkt{\frac{1}{\Lct^2}}}, 
\end{align}
where $N_{\rm cut}$ is determined as $m_{N_{\rm cut}}\simeq \Lct$. 
The leading terms are $\cO(N_{\rm cut}\Lct^4)=\cO(R\Lct^5)$. 

The above cutoff-dependent terms can be cancelled by introducing some heavy 5D fields~$\Psi_i$ 
with masses~$M_i$ ($i=1,2,\cdots,N_{\rm h}$).\footnote{
Such heavy fields can be understood as the Pauli-Villars regulators, but we regard them as physical fields in this paper. 
} 
Their contributions to $\rho_{\rm eng}$ are calculated in the same way, and obtain the total energy density as
\begin{align}
 \rho_{\rm eng} &= \sum_{n=0}^{N_{\rm cut}}\left[\frac{\Lct^4}{16\pi^2}\brkt{1+\sum_{i=1}^{N_{\rm h}}c_i}
 +\frac{\Lct^2}{16\pi^2}\brkt{m_n^2+\sum_{i=1}^{N_{\rm h}}c_im_n^{(i)2}}
 +\frac{1}{128\pi^2}\brkt{m_n^4+\sum_{i=1}^{N_{\rm h}}c_im_n^{(i)4}}
 \right.\nonumber\\
 &\hspace{13mm}\left.
 -\frac{m_n^4}{32\pi^2}\ln\frac{2\Lct}{m_n}-\sum_{i=1}^{N_{\rm h}}c_i\frac{m_n^{(i)4}}{32\pi^2}\ln\frac{2\Lct}{m_n^{(i)}}+\cdots\right], \nonumber\\
 p_{\rm prs} &= \sum_{n=0}^{N_{\rm cut}}\left[\frac{\Lct^4}{48\pi^2}\brkt{1+\sum_{i=1}^{N_{\rm h}}c_i}
 -\frac{\Lct^2}{48\pi^2}\brkt{m_n^2+\sum_{i=1}^{N_{\rm h}}c_im_n^{(i)2}}
 -\frac{7}{384\pi^2}\brkt{m_n^4+\sum_{i=1}^{N_{\rm h}}c_im_n^{(i)4}} \right.\nonumber\\
 &\hspace{13mm}\left.
 +\frac{m_n^4}{32\pi^2}\ln\frac{2\Lct}{m_n}+\sum_{i=1}^{N_{\rm h}}c_i\frac{m_n^{(i)4}}{32\pi^2}\ln\frac{2\Lct}{m_n^{(i)}}+\cdots\right], 
 \label{expr:rho-P}
\end{align}
where the ellipses denote terms suppressed by $\Lct$, and 
\begin{align}
 m_n^{(i)} &\equiv \sqrt{M_i^2+\frac{n^2}{R^2}}
\end{align}
are the KK masses. 
The integer~$c_i$ ($i=1,2,\cdots,N_{\rm h}$) denotes the degree of freedom and the statistics for the 5D field~$\Psi_i$. 
For example, $c_i=1$ for a scalar and $c_i=-4$ for a spinor. 
If we choose $c_i$ and $M_i$ to satisfy~\cite{Visser:2016mtr,Matsui:2024yoy} 
\begin{align}
 1+\sum_{i=1}^{N_{\rm h}}c_i &= 0, \;\;\;\;\;
 M^2+\sum_{i=1}^{N_{\rm h}}c_iM_i^2 = 0, \;\;\;\;\;
 M^4+\sum_{i=1}^{N_{\rm h}}c_iM_i^4 = 0, 
 \label{Pauli_cond}
\end{align}
\eqref{expr:rho-P} reduces to 
\begin{align}
 \rho_{\rm eng} &= -p_{\rm prs} = -\frac{m_n^4}{32\pi^2}\ln\frac{2\Lct}{m_n}-\sum_{i=1}^{N_{\rm h}}c_i\frac{m_n^{(i)4}}{32\pi^2}\ln\frac{2\Lct}{m_n^{(i)}}+\cdots 
 \nonumber\\
 &= \frac{m_n^4}{32\pi^2}\ln\frac{m_n}{\mu}+\sum_{i=1}^{N_{\rm h}}c_i\frac{m_n^{(i)4}}{32\pi^2}\ln\frac{m_n^{(i)}}{\mu}+\cdots, 
 \label{finite:rho_eng}
\end{align}
where $\mu$ is a renormalization scale that can be chosen to an arbitrary value due to the last condition in \eqref{Pauli_cond}.\footnote{ 
Thus, by tuning $\mu$, the second term in \eqref{finite:rho_eng} can be eliminated if desired. 
}
As a result, \eqref{finite:rho_eng} is expressed as
\begin{align}
 \rho_{\rm eng} &= -p_{\rm prs} = \frac{\mu^4}{64\pi^2}\sum_{n=0}^{N_{\rm cut}}F(\dlt_{\rm h}n), 
 \label{fin:rho_eng}
\end{align}
where $\dlt_{\rm h}\equiv (\mu R)^{-1}$, and
\begin{align}
 F(x) &\equiv \brkt{\hat{M}^2+x^2}^2\ln\brkt{\hat{M}^2+x^2}
 +\sum_{i=1}^{N_{\rm h}}c_i\brkt{\hat{M}_i^2+x^2}^2\ln\brkt{\hat{M}_i^2+x^2}, 
\end{align}
with
\begin{align}
 \hat{M} &\equiv \frac{M}{\mu}, \;\;\;\;\;
 \hat{M}_i \equiv \frac{M_i}{\mu}. 
\end{align}
The function~$F(x)$ damps to zero around $x=1$~\cite{Matsui:2024yoy}. 
In most of the literature, it is implicitly required that 
\begin{align}
 \lim_{R\to\infty}\Lvev{T_{MN}}=0, 
 \label{requirement}
\end{align}
where $\Lvev{T_{MN}}$ is the VEV of 5D the energy-momentum tensor. 
This indicates that $\rho_{\rm eng}$ and $p_{\rm prs}$ should be redefined as 
\begin{align}
 \frac{\hat{\rho}_{\rm eng}(R)}{\pi R} &\equiv \frac{\rho_{\rm eng}(R)}{\pi R}-\lim_{R\to\infty}\frac{\rho_{\rm eng}(R)}{\pi R}, \nonumber\\
 \frac{\hat{p}_{\rm prs}(R)}{\pi R} &\equiv \frac{p_{\rm prs}(R)}{\pi R}-\lim_{R\to\infty}\frac{p_{\rm prs}(R)}{\pi R}. 
 \label{redef:rhop}
\end{align}
Then we obtain the finite results.\footnote{
The derivation of the result from \eqref{fin:rho_eng} is shown in Ref.~\cite{Matsui:2024yoy}. 
} 
\begin{align}
 \hat{\rho}_{\rm eng} &= -\hat{p}_{\rm prs} = \cE_{\rm cas}(M)+\sum_{i=1}^{N_{\rm h}}c_i\cE(M_i), 
\end{align}
where
\begin{align}
 \cE_{\rm cas}(M) &\equiv -\frac{M^{\frac{5}{2}}}{16\pi^4R^{\frac{3}{2}}}\sum_{n=1}^\infty n^{-\frac{5}{2}}K_{\frac{5}{2}}(2\pi nRM), 
 \label{def:cE_cas}
\end{align}
where $K_\nu(z)$ is the modified Bessel function of the second kind. 
Since $\cE_{\rm cas}(M)$ becomes exponentially small for large values of $M$, 
all the contributions of the heavy fields~$\Psi_i$ can be neglected. 
Using some mathematical formulae, we can rewrite \eqref{def:cE_cas} as
\begin{align}
 \cE_{\rm cas}(M) &= \frac{M^5R}{4\pi}\sbk{\frac{1}{\alp^3}{\rm Li}_3(e^{-\alp})+\frac{3}{\alp^4}{\rm Li}_4(e^{-\alp})+\frac{3}{\alp^5}{\rm Li}_5(e^{-\alp})}, 
 \label{cv_result:massive}
\end{align}
where $\alp\equiv 2\pi RM$. 
In particular, for the massless scalar ($M=0$), we obtain
\begin{align}
 \hat{\rho}_{\rm eng} &= -\hat{p}_{\rm prs} = \cE_{\rm cas}(0) = -\frac{3\zeta(5)}{128\pi^6R^4}. 
 \label{cv_result:massless}
\end{align}
This is the result that agrees with the one obtained by the regularizations based on the analytic continuation, 
such as the dimensional regularization. 
However, the physical meaning of the requirement~\eqref{requirement} is unclear. 
Since the redefinition of the energy density and the pressure in \eqref{redef:rhop} does not affect their $R$-dependence, 
it causes no problems when we discuss the modulus stabilization. 
In contrast, when we discuss the time evolution of the bulk space, the absolute value of $\Lvev{T_{MN}}$ is relevant, 
and thus we have to justify the modifications in \eqref{redef:rhop} if we adopt them.  
Hence we do not require \eqref{requirement} in the following. 

In this paper, we assume that all heavy masses~$M_i$ are the same order, i.e., $M_i=\cO(M_{\rm h})$, 
and $M\ll M_{\rm h}< \Lct$. 
In the case that the conditions in \eqref{Pauli_cond} are satisfied,\footnote{
As an example that satisfies \eqref{Pauli_cond}, we can consider one light scalar with mass~$M$, 
seven heavy scalars thet have a common mass~$M_{\rm B}$ and two heavy Dirac spinors whose masses are $M_{\rm F1}$ and $M_{\rm F2}$. 
If the masses are related as
\begin{align}
 M_{\rm B}^2 &= \frac{28-4\sqrt{7}}{21}M_{\rm F2}^2+\frac{4\sqrt{7}-7}{21}M^2, \;\;\;\;\;
 M_{\rm F1}^2 = \frac{4-\sqrt{7}}{3}M_{\rm F2}^2+\frac{\sqrt{7}-1}{3}M^2, 
\end{align}
the all conditions in \eqref{Pauli_cond} are satisfied. 
}  
we can safely take a limit~$\Lct\to\infty$, and
now $M_{\rm h}$ is understood as a cutoff scale.
However, we will discuss a possibility that \eqref{Pauli_cond} is not satisfied in the following sections. 
Thus, we keep $\Lct$ finite as the energy scale at which the 5D theory is replaced with a more fundamental theory. 
Namely, we regard it as a physical scale such as a typical mass scale of some new particles or the string excitation modes, 
rather than an artificial regularization parameter. 

Here we comment on other regularization schemes. 
The regularization schemes based on the analytic continuation,  
such as the dimensional regularization and the zeta-function regularization, are more efficient to calculate the moduli-dependent part 
of the vacuum energy, i.e., the Casimir energy. 
However, such schemes cannot deal with the divergences corresponding to the positive power of $\Lct$ 
(see Sec.~2 of Ref.~\cite{Matsui:2024yoy}, for example). 
They essentially just drop such divergences, and see only the logarithmic divergence. 
More precisely, we need to analyze the pole structures other than that at $\ep=0$, 
where $\ep$ parameterizes the deviation from the physical dimension, 
in order to see the positive-power divergent terms in $\Lct$~\cite{Visser:2016mtr}. 
Besides, the relation between $\ep$ and the energy scale~$\Lct$ is unclear in these regularizations.  
Hence, such regularization schemes are not appropriate for our purpose.

\section{5D energy-momentum tensor}
\label{5D_EMT}
When we consider the cosmological evolution of the spacetime with a large extra dimension, such as the dark dimension scenario, 
we need to solve the 5D field equations because the temperature of the universe can be higher than the KK scale, 
which is the cutoff scale of the 4D effective theory. 
The evolution of the spacetime is determined by the Einstein equation,\footnote{
We adopt the semiclassical approach~\cite{Birrell:1982ix}, in which the spacetime is treated classically, while the quantum effects of matter fields 
are incorporated via $\Lvev{T_{MN}}$. 
} 
\begin{align}
 \cR_{MN}-\frac{1}{2}g_{MN}\cR+g_{MN}\Lmd_{\rm cc}^{\rm (5D)} &= \kp_5\Lvev{T_{MN}}, 
 \label{Einstein_eq}
\end{align}
where $\cR_{MN}$ and $\cR\equiv g^{MN}\cR_{MN}$ are the 5D Ricci tensor and scalar, respectively, 
$\Lmd_{\rm cc}^{\rm (5D)}$ is the 5D cosmological constant, and 
$\kp_5$ is the 5D gravitational coupling, 
and $\Lvev{T_{MN}}$ is the expectation value of the energy-momentum tensor~$T_{MN}$ for quantum fields. 
In the following, we calculate $\Lvev{T_{MN}}$ for scalar and spinor fields. 

Since the background space expands slowly enough compared to the quantum time scale, 
we can neglect the time-dependence of the background metric 
in the calculation of $\Lvev{T_{MN}}$. 
Thus, in the following calculations, the 5D metric is parameterized as 
\begin{align}
 ds^2 &= g_{MN}dx^Mdx^N = -n^2(y)dt^2+a^2(y)\sum_{i=1}^3(dx^i)^2+b^2(y)dy^2, 
 \label{metric_ansatz}
\end{align}
where $t\equiv x^0$, $y\equiv x^4$, and $0\leq y\leq L\equiv \pi R$. 
To simplify the calculations, we further assume that 
\begin{align}
 a(y) &= \asc n(y), 
 \label{metric_assumption}
\end{align}
where $\asc$ is a constant. 
Although $\asc$ can always be set to 1 by the redefinition of the 3D coordinates~$x^i$ ($i=1,2,3$), 
we leave it as an arbitrary positive constant because it becomes the scale factor when we revive the time-dependence of the background metric. 
In fact, when we solve the evolution equation~\eqref{Einstein_eq}, 
we should revive the time-dependence of $n$, $\asc$ and $b$ in the our results~\eqref{expr:T_MN:boson1}, \eqref{expr:T_MN:boson2} and \eqref{expr:T_MN:fermion}.  
The explicit forms of the evolution equations are given by \eqref{evolve_eq:restrict} with \eqref{constraint:restrict1} and \eqref{constraint:restrict2} 
in Appendix~\ref{Einstein_eqs:2}.

\subsection{Bosonic contribution} \label{bosonic}
In this subsection, we calculate the contribution from a real scalar~$\Phi(x^\mu,y)$, whose Lagrangian is given by \eqref{cL:scalar}. 
Since $\Phi$ is $Z_2$-even, the boundary conditions are 
\begin{align}
 \der_y\Phi(x^\mu,0) &= 0, \;\;\;\;\;
 \der_y\Phi(x^\mu,L) = 0, 
 \label{BC:scalar}
\end{align}
where $x^\mu$ denotes the 4D coordinates.

\subsubsection{Equation of motion}
The equation of motion is 
\begin{align}
 \frac{1}{\sqrt{-g}}\der_M\brkt{\sqrt{-g}g^{MN}\der_N\Phi}-M^2\Phi &= 0. 
\end{align}
Here we move to the momentum basis for the 4D coordinates~$x^\mu$. 
Then, under our metric ansatz, the above equation is expressed as
\begin{align}
 \brc{\frac{\pf^2}{n^2}-\frac{1}{b^2}\sbk{\der_y^2+\brkt{\frac{4n'}{n}-\frac{b'}{b}}\der_y}+M^2}\tl{\Phi} &= 0, 
 \label{EOM:tlPhi}
\end{align}
where the prime denotes the $y$-derivative, 
\begin{align}
 \pf^2 &\equiv -p_t^2+\frac{1}{\asc^2}\sum_{i=1}^3p_i^2, 
\end{align}
and $\tl{\Phi}(p_\mu,y)$ is defined as 
\begin{align}
 \Phi(x^\mu,y) &= \int\frac{d^4p}{(2\pi)^4}\;e^{ip\cdot x}\tl{\Phi}(p_\mu,y). \;\;\;\;\; \brkt{p\cdot x\equiv p_\mu x^\mu}
\end{align}

\subsubsection{5D propagator}
Following the procedure in Ref.~\cite{Gherghetta:2000kr}, 
we consider the 5D propagator, 
\begin{align}
 \langle 0|T\Phi(x^\mu,y)\Phi(x^{\prime \nu},y')|0\rangle 
 &\equiv \int \frac{d^4p}{i(2\pi)^4}\;e^{ip\cdot (x-x')}\tl{G}_{\rm B}(\pf^2,y,y'). 
\end{align}
As shown in App.~\ref{5D_propagator:scalar_sector}, 
the Fourier component~$\tl{G}_{\rm B}(\pf^2,y,y')$ is expressed as
\begin{align}
 \tl{G}_{\rm B}(\pf^2,y,y') &= \vth(y-y')\tl{G}_{\rm B>}(\pf^2,y,y')+\vth(y'-y)\tl{G}_{\rm B<}(\pf^2,y,y'), 
\end{align}
where $\vth(y)$ is the Heaviside step function, and 
\begin{align}
 \tl{G}_{\rm B>}(\pf^2,y,y') &= \frac{b(L)}{\asc^3n^4(L)}\frac{C_L(y;\pf^2)C_0(y';\pf^2)}{C_0'(L;\pf^2)}, \nonumber\\
 \tl{G}_{\rm B<}(\pf^2,y,y') &= \frac{b(L)}{\asc^3n^4(L)}\frac{C_0(y;\pf^2)C_L(y';\pf^2)}{C_0'(L;\pf^2)}. 
\end{align}
The functions~$C_{0,L}(y;\pf^2)$ are the basis of solutions to \eqref{EOM:tlPhi}, 
and their definitions are provided in Appendix~\ref{scalar:basis_fct}.

\subsubsection{Energy-momentum tensor}
The energy momentum tensor is 
\begin{align}
 T_{MN}^{\rm b} &= \der_M\Phi\der_N\Phi-g_{MN}\brkt{\frac{1}{2}g^{PQ}\der_P\Phi\der_Q\Phi+\frac{M^2}{2}\Phi^2} \nonumber\\
 &= \lim_{x'\to x}\lim_{y'\to y}\sbk{\der_M\der'_N-g_{MN}\brkt{\frac{1}{2}g^{PQ}\der_P\der'_Q+\frac{M^2}{2}}}\Phi(x^\mu,y)\Phi(x^{\prime \nu},y'). 
\end{align}
Thus, the expectation values of its components are expressed as
\begin{align}
 \Lvev{T^{\rm b}_{tt}} &= \lim_{x'\to x}\lim_{y'\to y}
 \sbk{\der_t\der'_t-g_{tt}\brkt{\frac{1}{2}g^{\rho\sgm}\der_\rho\der'_\sgm+\frac{1}{2}g^{yy}\der_y\der'_y+\frac{M^2}{2}}} 
 \int \frac{d^4p}{i(2\pi)^4}\;e^{ip\cdot (x-x')}\tl{G}_{\rm B}(\pf^2,y,y')  \nonumber\\
 &= \lim_{y'\to y}\int\frac{d^4p}{2i(2\pi)^4}\;\sbk{p_t^2+\frac{\pth^2}{\asc^2}+\frac{n^2(y)}{b^2(y)}\der_y\der'_y+n^2(y)M^2}
 \tl{G}_{\rm B}(\pf^2,y,y'), \nonumber\\
 \Lvev{T^{\rm b}_{ij}} &= \lim_{y'\to y}\int\frac{d^4p}{i(2\pi)^4}\;
 \sbk{p_ip_j-g_{ij}\brkt{\frac{\pf^2}{2n^2(y)}+\frac{1}{2b^2(y)}\der_y\der'_y+\frac{M^2}{2}}}\tl{G}_{\rm B}(\pf^2,y,y') \nonumber\\
 &= \asc^2\dlt_{ij}\lim_{y'\to y}\int\frac{d^4p}{2i(2\pi)^4}\;
 \sbk{p_t^2-\frac{\pth^2}{3\asc^2}-\frac{n^2(y)}{b^2(y)}\der_y\der'_y-n^2(y)M^2}\tl{G}_{\rm B}(\pf^2,y,y'), 
 \nonumber\\
 \Lvev{T^{\rm b}_{yy}} 
 &= \lim_{y'\to y}\int\frac{d^4p}{2i(2\pi)^4}\;
 \sbk{\frac{b^2(y)}{n^2(y)}\brkt{p_t^2-\frac{\pth^2}{\asc^2}}+\der_y\der'_y-b^2(y)M^2}\tl{G}_{\rm B}(\pf^2,y,y'), 
\end{align}
where $i,j=1,2,3$, and 
\begin{align}
 \pth^2 &\equiv p_1^2+p_2^2+p_3^2. 
\end{align}
We have used that
\begin{align}
 \int d^4p\;p_ip_j\cF(p_t,\pth^2) &= \frac{\dlt_{ij}}{3}\int d^4p\;\pth^2\cF(p_t,\pth^2), 
 \label{odd_int}
\end{align}
for an arbitrary function~$\cF(p_t,\pth^2)$. 
The expectation values of the off-diagonal components vanish 
because integrals of an odd power of $p_\mu$ vanish. 

Now we perform the Wick rotation~$p_t\to i\tl{p}_t$.\footnote{
Since we have neglected the $t$-dependence of the background geometry, we can perform the Wick rotation without encountering 
any problems discussed in Ref.~\cite{Visser:2017atf}. 
}
Then, we have
\begin{align}
 \Lvev{T^{\rm b}_{tt}} &= \lim_{y'\to y}\int_{-\infty}^\infty\frac{d\tl{p}_t}{4\pi}\int\frac{d^3p}{(2\pi)^3}\;
 \sbk{-\tl{p}_t^2+\frac{\pth^2}{\asc^2}+\frac{n^2(y)}{b^2(y)}\der_y\der_y'+n^2(y)M^2}\tl{G}_{\rm B}(\tl{p}_4^2,y,y') \nonumber\\
 &= \frac{\asc^3}{4\pi^3}\lim_{y'\to y}\int_0^\infty d\tl{p}_t\int_0^\infty d\tl{p}_{\rm 3d}\;\tl{p}_{\rm 3d}^2
 \sbk{-\tl{p}_t^2+\tpt^2+\frac{n^2(y)}{b^2(y)}\der_y\der_y'+n^2(y)M^2}\tl{G}_{\rm B}(\tl{p}_4^2,y,y'), 
\end{align}
where $\tpt\equiv \pth/\asc$ and $\tl{p}_4^2\equiv \tl{p}_t^2+\tpt^2$. 
Here we change the integration variables as $(\tl{p}_t,\tpt)\to (\rho,\tht)$, where 
\begin{align}
 \tl{p}_t &\equiv \rho\cos\tht, \;\;\;\;\;
 \tpt \equiv \rho\sin\tht. 
 \label{change:int_var}
\end{align}
Then, the above expression can be rewritten as
\begin{align}
 \Lvev{T^{\rm b}_{tt}} &= \frac{\asc^3}{4\pi^3}\lim_{y'\to y}\int_0^\infty d\rho\int_0^{\pi/2}d\tht\; \rho^3\sin^2\tht \nonumber\\
 &\hspace{40mm}
 \times\sbk{\rho^2\brkt{-\cos^2\tht+\sin^2\tht}+\frac{n^2(y)}{b^2(y)}\der_y\der_y'+n^2(y)M^2}\tl{G}_{\rm B}(\rho^2,y,y') \nonumber\\
 &=\frac{\asc^3}{16\pi^2}\lim_{y'\to y}\int_0^\infty d\rho\;\rho^3\sbk{
 \frac{\rho^2}{2}+\frac{n^2(y)}{b^2(y)}\der_y\der_y'+n^2(y)M^2}
 \tl{G}_{\rm B}(\rho^2,y,y'). 
 \label{expr:T_MN:boson1}
\end{align}
Similarly, we have 
\begin{align}
 \Lvev{T^{\rm b}_{ij}} 
 &= -\frac{\asc^5\dlt_{ij}}{16\pi^2}\lim_{y'\to y}\int_0^\infty d\rho\;\rho^3
 \sbk{\frac{\rho^2}{2}+\frac{n^2(y)}{b^2(y)}\der_y\der_y'+n^2(y)M^2}\tl{G}_{\rm B}(\rho^2,y,y'), \nonumber\\
 \Lvev{T^{\rm b}_{yy}} 
 &= -\frac{\asc^3b^2(y)}{16\pi^2n^2(y)}\lim_{y'\to y}\int_0^\infty d\rho\;\rho^3
 \sbk{\rho^2-\frac{n^2(y)}{b^2(y)}\der_y\der_y'+n^2(y)M^2}\tl{G}_{\rm B}(\rho^2,y,y'). 
 \label{expr:T_MN:boson2}
\end{align}

\subsection{Fermionic contribution}
\subsubsection{Equations of motion}
Here we calculate the contribution from a 5D spinor field~$\Psi$, whose Lagrangian is given by
\begin{align}
 \cL_{\rm spinor} &= i\bar{\Psi}\Gm^A E_A^{\;\;M}\cD_M\Psi+M\bar{\Psi}\Psi, 
\end{align}
where $E_A^{\;\;M}$ is the inverse vielbein, $A=0,1,\cdots,4$ denote the flat indices, $\Gm^A$ are the 5D gamma matrices, and 
\begin{align}
 \cD_M &\equiv \der_M-\frac{1}{8}\omg_M^{\;\;AB}\sbk{\Gm_A,\Gm_B}. 
\end{align}
The spin connection is given by
\begin{align}
 \omg_M^{\;\;AB} &= E_N^{\;\;A}\brkt{\der_M E^{NB}+\Gm^N_{\;\;LM}E^{LB}}, 
\end{align}
where $\Gm^N_{\;\;LM}$ is the Christoffel symbol. 

In terms of the 2-component spinors~$\chi$ and $\bar{\lmd}$, defined by 
\begin{align}
 \Psi &= \begin{pmatrix} \chi \\ \bar{\lmd} \end{pmatrix}, 
\end{align}
the Dirac equation is expressed as
\begin{align}
 -\frac{i}{n}\brkt{\der_t\bar{\lmd}-\frac{in'}{2b}\chi}+\frac{i}{a}\brkt{\tau^i\der_i\bar{\lmd}+\frac{3ia'}{2b}\chi}
 -\frac{1}{b}\der_y\chi+M\chi &= 0, \nonumber\\
 -\frac{i}{n}\brkt{\der_t\chi+\frac{in'}{2b}\bar{\lmd}}-\frac{i}{a}\brkt{\tau^i\der_i\chi+\frac{3ia'}{2b}\bar{\lmd}}
 +\frac{1}{b}\der_y\bar{\lmd}+M\bar{\lmd} &= 0, 
\end{align}
where $\tau^i$ ($i=1,2,3$) are the Pauli matrices. 
Moving to the 4D momentum basis, and rescaling the fields as
\begin{align}
 \hat{\chi}(p_\mu,y) &\equiv n^2(y)\tl{\chi}(p_\mu,y), \;\;\;\;\;
 \bar{\hat{\lmd}}(p_\mu,y) \equiv n^2(y)\bar{\tl{\lmd}}(p_\mu,y), 
\end{align}
where
\begin{align}
 \chi(x^\mu,y) &\equiv \int\frac{d^4p}{(2\pi)^4}\;e^{ip\cdot x}\tl{\chi}(p_\mu,y), \;\;\;\;\;
 \bar{\lmd}(x^\mu,y) \equiv \int\frac{d^4p}{(2\pi)^4}\;e^{ip\cdot x}\bar{\tl{\lmd}}(p_\mu,y), 
\end{align}
the Dirac equation becomes 
\begin{align}
 \frac{1}{n}\brkt{p_t-\frac{1}{\asc}\tau^ip_i}\bar{\hat{\lmd}}-\frac{1}{b}\brkt{\der_y-bM}\hat{\chi} &= 0, \nonumber\\
 \frac{1}{n}\brkt{p_t+\frac{1}{\asc}\tau^ip_i}\hat{\chi}+\frac{1}{b}\brkt{\der_y+bM}\bar{\hat{\lmd}} &= 0. 
 \label{Dirac_eq}
\end{align}

Here we choose the boundary conditions for $\chi$ as
\begin{align}
 \hat{\chi}(p_\mu,0) &= \hat{\chi}(p_\mu,L) = 0. 
 \label{BC:chi}
\end{align}
Then, from \eqref{Dirac_eq},  $\bar{\hat{\lmd}}$ is subject to 
\begin{align}
 \left.\brkt{\der_y+bM}\bar{\hat{\lmd}}\right|_{y=0} &= \left.\brkt{\der_y+bM}\bar{\hat{\lmd}}\right|_{y=L} = 0. 
 \label{BC:lmd}
\end{align} 

From \eqref{Dirac_eq}, we obtain
\begin{align}
 \cO_-\hat{\chi} &= 0, \;\;\;\;\;
 \cO_+\bar{\hat{\lmd}} = 0, 
\end{align}
where
\begin{align}
 \cO_\pm &\equiv \frac{b^2}{n^2}\pf^2
 -\der_y^2-\brkt{\frac{n'}{n}-\frac{b'}{b}}\der_y \mp\frac{n'}{n}bM+b^2M^2. 
 \label{def:cO_pm}
\end{align}

\subsubsection{5D propagator}
The 5D propagator 
\begin{align}
 \langle 0|T\Psi(x^\mu,y)\bar{\Psi}(x^{\prime \nu},y')|0\rangle 
 &= \int \frac{d^4p}{i(2\pi)^4}\;e^{ip\cdot(x-x')}\frac{\hat{G}_{\rm F}(p_\mu,y,y')}{n^2(y)n^2(y')}, 
\end{align}
is expressed in the 2-component spinor notation as
\begin{align}
 \hat{G}_{\rm F}(p_\mu,y,y') &= -\begin{pmatrix} \hat{G}_{\chi\lmd}(p_\mu,y,y') & \hat{G}_{\chi\chi}(p_\mu,y,y') \\
 \hat{G}_{\lmd\lmd}(p_\mu,y,y') & \hat{G}_{\lmd\chi}(p_\mu,y,y') \end{pmatrix}. 
\end{align}
As shown in App.~\ref{5Dprop:spinor}, the components of the 5D propagator have the following forms. 
\begin{align}
 \hat{G}_{\chi\chi}(p_\mu,y,y') &= \bar{G}_{\chi\chi}(\pf^2,y,y')\brkt{p_t\id_2-\frac{p_i}{\asc}\tau^i}, \nonumber\\
 \hat{G}_{\chi\lmd}(p_\mu,y,y') &= \bar{G}_{\chi\lmd}(\pf^2,y,y')\id_2, \nonumber\\
 \hat{G}_{\lmd\chi}(p_\mu,y,y') &= \bar{G}_{\lmd\chi}(\pf^2,y,y')\id_2, \nonumber\\
 \hat{G}_{\lmd\lmd}(p_\mu,y,y') &= \bar{G}_{\lmd\lmd}(\pf^2,y,y')\brkt{p_t\id_2+\frac{p_i}{\asc}\tau^i}, 
\end{align}
where the functions~$\bar{G}_{ab}(\pf^2,y,y')$ ($a,b=\chi,\lmd$) are read off from \eqref{expr:hatGs}, 
which are expressed in terms of the basis functions for $\cO_\pm$ defined in Appendix~\ref{spinor:basis_fct}.

\subsubsection{Energy-momentum tensor}
The energy-momentum tensor is
\begin{align}
 T^{\rm f}_{MN} &= -\frac{i}{4}\brkt{\bar{\Psi}\Gm_N\cD_M\Psi-\cD_M\bar{\Psi}\Gm_N\Psi}+\brkt{M\leftrightarrow N} \nonumber\\
 &= \frac{i}{4}{\rm Tr}\brkt{\Gm_N\cD_M\Psi\bar{\Psi}-\Gm_N\Psi\cD_M\bar{\Psi}}+\brkt{M\leftrightarrow N} \nonumber\\
 &= \frac{i}{4}\lim_{x'\to x}\lim_{y'\to y}{\rm Tr}\brc{\Gm_N\brkt{\der_M-\der_M'-\frac{1}{4}\omg_M^{\;\;AB}\sbk{\Gm_A,\Gm_B}}\Psi(x^\mu,y)\bar{\Psi}(x^{\prime\nu},y')}
 +\brkt{M\leftrightarrow N}, 
\end{align}
where $\Gm_N\equiv E_N^{\;\;C}\Gm_C$, and $\der_M'\equiv \der/\der x^{\prime M}$. 
Thus, the expectation values~$\Lvev{T^{\rm f}_{MN}}$ is expressed 
in terms of the 5D propagator~$\hat{G}_{\rm F}(p_\mu,y,y')$ as
\begin{align}
 \Lvev{T^{\rm f}_{\mu\nu}} &= \frac{i}{4}\lim_{x'\to x}\lim_{y'\to y}{\rm Tr}
 \brc{\Gm_\nu\brkt{\der_\mu-\der_\mu'-\frac{1}{4}\omg_\mu^{\;\;AB}\sbk{\Gm_A,\Gm_B}}
 \int\frac{d^4p}{i(2\pi)^4}\;e^{ip\cdot (x-x')}\frac{\hat{G}_{\rm F}(p_\mu,y,y')}{n^2(y)n^2(y')}} \nonumber\\
 &\quad
 +\brkt{\mu\leftrightarrow \nu} \nonumber\\
 &= \frac{i}{4}\lim_{y'\to y}\int\frac{d^4p}{i(2\pi)^4}\;{\rm Tr}\brc{\Gm_\nu\brkt{2ip_\mu-\frac{1}{4}\omg_\mu^{\;\;AB}\sbk{\Gm_A,\Gm_B}}
 \frac{\hat{G}_{\rm F}(p_\mu,y,y')}{n^2(y)n^2(y')}}+\brkt{\mu\leftrightarrow\nu}, \nonumber\\
 \Lvev{T^{\rm f}_{\mu y}} &= \frac{i}{4}\lim_{y'\to y}\int\frac{d^4p}{i(2\pi)^4}\;{\rm Tr}
 \brc{\Gm_y\brkt{2ip_\mu-\frac{1}{4}\omg_\mu^{\;\;AB}\sbk{\Gm_A,\Gm_B}}\frac{\hat{G}_{\rm F}(p_\mu,y,y')}{n^2(y)n^2(y')}} \nonumber\\
 &\quad
 +\frac{i}{4}\lim_{y'\to y}\int\frac{d^4p}{i(2\pi)^4}\;{\rm Tr}
 \brc{\Gm_\mu\brkt{\der_y-\der_y'-\frac{1}{4}\omg_y^{\;\;AB}\sbk{\Gm_A,\Gm_B}}\frac{\hat{G}_{\rm F}(p_\mu,y,y')}{n^2(y)n^2(y')}}, \nonumber\\
 \Lvev{T^{\rm f}_{yy}} &= \frac{i}{2}\lim_{y'\to y}\int\frac{d^4p}{i(2\pi)^4}\;{\rm Tr}
 \brc{\Gm_y\brkt{\der_y-\der_y'-\frac{1}{4}\omg_y^{\;\;AB}\sbk{\Gm_A,\Gm_B}}\frac{\hat{G}_{\rm F}(p_\mu,y,y')}{n^2(y)n^2(y')}}. 
\end{align}
Since 
\begin{align}
 \omg_t^{\;\;AB}\sbk{\Gm_A,\Gm_B} &= 2\omg_t^{\;\;04}\sbk{\Gm_0,\Gm_4} = \frac{4n'}{b}\begin{pmatrix} & -i\id_2 \\ i\id_2 & \end{pmatrix}, \nonumber\\
 \omg_i^{\;\;AB}\sbk{\Gm_A,\Gm_B} &= 2\omg_i^{\;\;\udl{j}4}\sbk{\Gm_{\udl{j}},\Gm_4} = \frac{4a'}{b}\dlt_{ij}
 \begin{pmatrix} & -i\tau^j \\ -i\tau^j & \end{pmatrix}, \nonumber\\
 \omg_y^{\;\;AB}\sbk{\Gm_A,\Gm_B} &= \bdm{0}_4, 
\end{align}
where $\tau^j$ ($j=1,2,3$) are the Pauli matrices, 
we have the following expressions. 
\begin{align}
 \Lvev{T^{\rm f}_{tt}} &= \frac{1}{n^3(y)}\lim_{y'\to y}\int\frac{d^4p}{i(2\pi)^4}\;p_t{\rm Tr}\sbk{\hat{G}_{\chi\chi}(p_\mu,y,y')+\hat{G}_{\lmd\lmd}(p_\mu,y,y')} \nonumber\\
 &= \frac{2}{n^3(y)}\lim_{y'\to y}\int\frac{d^4p}{i(2\pi)^4}\;p_t^2\sbk{\bar{G}_{\chi\chi}(\pf^2,y,y')+\bar{G}_{\lmd\lmd}(\pf^2,y,y')}, 
 \nonumber\\
 \Lvev{T^{\rm f}_{ij}} &= \frac{a(y)}{2n^4(y)}\lim_{y'\to y}\int\frac{d^4p}{i(2\pi)^4}\;{\rm Tr}\sbk{\brkt{p_i\tau^j+p_j\tau^i}
 \brkt{\hat{G}_{\lmd\lmd}(p_\mu,y,y')-\hat{G}_{\chi\chi}(p_\mu,y,y')}} \nonumber\\
 &= \frac{2\dlt_{ij}}{3n^3(y)}\lim_{y'\to y}\int\frac{d^4p}{i(2\pi)^4}\;\pth^2\sbk{\bar{G}_{\chi\chi}(\pf^2,y,y')+\bar{G}_{\lmd\lmd}(\pf^2,y,y')}, 
 \nonumber\\
 \Lvev{T^{\rm f}_{yy}} &= \frac{b(y)}{2}\lim_{y'\to y}\int\frac{d^4p}{i(2\pi)^4}\;{\rm Tr}
 \sbk{\brkt{\der_y-\der_y'}\brkt{\frac{\hat{G}_{\chi\lmd}(p_\mu,y,y')}{n^2(y)n^2(y')}-\frac{\hat{G}_{\lmd\chi}(p_\mu,y,y')}{n^2(y)n^2(y')}}} \nonumber\\
 &= b(y)\lim_{y'\to y}\int\frac{d^4p}{i(2\pi)^4}\;
 \sbk{\brkt{\der_y-\der_y'}\brkt{\frac{\bar{G}_{\chi\lmd}(\pf^2,y,y')}{n^2(y)n^2(y')}-\frac{\bar{G}_{\lmd\chi}(\pf^2,y,y')}{n^2(y)n^2(y')}}}. 
\end{align}
We have used \eqref{odd_int}. 
The expectation values of the off-diagonal components are zero 
because integrals of an odd power of $p_\mu$ vanish. 

Now we perform the Wick rotation and the change of integration variables in \eqref{change:int_var}, and obtain the following expressions. 
\begin{align}
 \Lvev{T^{\rm f}_{tt}} 
 &= -\frac{\asc^3}{16\pi^2n^3(y)}\lim_{y'\to y}\int_0^\infty d\rho\;\rho^5
 \sbk{\bar{G}_{\chi\chi}(\rho^2,y,y')+\bar{G}_{\lmd\lmd}(\rho^2,y,y')}, \nonumber\\
 \Lvev{T^{\rm f}_{ij}}
 &= \frac{\asc^5\dlt_{ij}}{16\pi^2n^3(y)}\lim_{y'\to y}\int_0^\infty d\rho\;\rho^5
 \sbk{\bar{G}_{\chi\chi}(\rho^2,y,y')+\bar{G}_{\lmd\lmd}(\rho^2,y,y')}, \nonumber\\
 \Lvev{T^{\rm f}_{yy}} 
 &= \frac{\asc^3b(y)}{8\pi^2n^4(y)}\lim_{y'\to y}\int_0^\infty d\rho\;\rho^3
 \sbk{\brkt{\der_y-\der_y'}\brkt{\bar{G}_{\chi\lmd}(\rho^2,y,y')-\bar{G}_{\lmd\chi}(\rho^2,y,y')}}. 
 \label{expr:T_MN:fermion}
\end{align}

\section{Cutoff-dependence} \label{cutoff_dep}
Note that the expressions in \eqref{expr:T_MN:boson1}, \eqref{expr:T_MN:boson2} and \eqref{expr:T_MN:fermion} diverges by taking the limit of $y'\to y$. 
Since we are interested in the dependence on the cutoff scale~$\Lct$, 
we keep $\Lct\equiv (y'-y)^{-1}$ finite.\footnote{ 
In our setup, we have neglected any interaction terms because they are irrelevant to the one-loop contributions to the vacuum energy density and the pressures. 
Thus the theory is renormalizable. 
However, once interactions are turned on, it becomes non-renormalizable and 
should be regarded as effective theory that is valid only below some cut-off energy scale~$\Lct$. \label{fn:renormalizability}
}
We should also notice that there is an ambiguity in a way of introducing $\Lct$ to the expressions of $\Lvev{T_{MN}}$. 
For example, \eqref{expr:T_MN:boson1} can be regularized as
\begin{align}
 \Lvev{T^{\rm b}_{tt}} &= \frac{\asc^3}{16\pi^2}\int_0^\infty d\rho\;\rho^3
 \sbk{\frac{\rho^2}{2}+\frac{n^2(y)}{b^2(y)}\der_y\der_y'+n^2(y)M^2}\tilde{G}_{\rm B}(\rho^2,y,y'), 
 \label{reg:T_tt:1}
\end{align}
or
\begin{align}
 \Lvev{T^{\rm b}_{tt}} &= \frac{\asc^3}{16\pi^2}\int_0^\infty d\rho\;\rho^3
 \sbk{\frac{\rho^2}{2}+\frac{n(y)n(y')}{b(y)b(y')}\der_y\der_y'+n(y)n(y')M^2}\tilde{G}_{\rm B}(\rho^2,y,y'), 
 \label{reg:T_tt:2}
\end{align}
or
\begin{align}
 \Lvev{T^{\rm b}_{tt}} &= \frac{\asc^3}{16\pi^2}\int_0^\infty d\rho\;\rho^3
 \sbk{\frac{\rho^2}{2}+\frac{n^2(\yh)}{b^2(\yh)}\der_y\der_y'+n^2(\yh)M^2}\tilde{G}_{\rm B}(\rho^2,y,y'), 
 \label{reg:T_tt:3}
\end{align}
where
\begin{align}
 y' &= y+\frac{1}{\Lct}, \;\;\;\;\;
 \yh \equiv \frac{y+y'}{2} = y+\frac{1}{2\Lct}. 
\end{align}
We have checked that this ambiguity does not essentially change our conclusion. 
Hence, in this paper, we choose $\yh$ as the arguments of the background metric components~$n$ and $b$ in front of the 5D propagators, 
just like \eqref{reg:T_tt:3}, in the following calculations.

\subsection{Scalar sector}
In order to extract the divergent part (i.e., terms with positive power of $\Lct$) of $\Lvev{T_{MN}^{\rm b}}$, we divide the $\rho$-integral as
\begin{align}
 \int_0^\infty d\rho &= \int_0^{\bar{\rho}}d\rho+\int_{\bar{\rho}}^\infty d\rho. 
 \label{divide:int}
\end{align}
and the corresponding parts to the first and the second terms are denoted as $\Lvev{T_{MN}^{\rm b}}^{\rm fin}$ 
and $\Lvev{T_{MN}^{\rm b}}^{\rm UV}$, respectively.  
Then, the divergent terms are included in $\Lvev{T_{tt}^{\rm b}}^{\rm UV}$. 
The constant~$\bar{\rho}$ is chosen to satisfy
\begin{align}
 \bar{\rho}\int_y^{y'}d\tl{y}\;\frac{b(\tl{y})}{n(\tl{y})} &\simeq \frac{\bar{\rho}}{\Lct}\frac{b(y)}{n(y)} \simgt \cO(1). 
\end{align}

The behavior of the 5D propagator at large $\rho$ is shown in Appendix~\ref{apx_expr:boson}. 
Using \eqref{tlG_B} and \eqref{der2tlG_B}, we have
\begin{align}
 &\quad
 \sbk{\frac{\rho^2}{2}+\frac{n^2(\yh)}{b^2(\yh)}\der_y\der_y'+n^2(\yh)M^2}\tl{G}_{\rm B<}(\rho^2,y,y') \nonumber\\
 &= \frac{e^{-\rho\Dlt(y,y')}}{2\asc^3n^{3/2}(y)n^{3/2}(y')}
 \sbk{\rho\cT_{-1}^{(tt)}+\cT_0^{(tt)}+\frac{\cT_1^{(tt)}}{\rho}+\frac{\cT_2^{(tt)}}{\rho^2}+\frac{\cT_3^{(tt)}}{\rho^3}+\cdots}, 
\end{align}
and
\begin{align}
 &\quad
 \sbk{\rho^2-\frac{n^2(\yh)}{b^2(\yh)}\der_y\der_y'+n^2(\yh)M^2}\tl{G}_{\rm B<}(\rho^2,y,y') \nonumber\\
 &= \frac{e^{-\rho\Dlt(y,y')}}{2\asc^3n^{3/2}(y)n^{3/2}(y')}
 \sbk{\rho\cT_{-1}^{(yy)}+\cT_0^{(yy)}+\frac{\cT_1^{(yy)}}{\rho}+\frac{\cT_2^{(yy)}}{\rho^2}+\frac{\cT_3^{(yy)}}{\rho^3}+\cdots}, 
\end{align}
where
\begin{align}
 \Dlt(y,y') &\equiv \int_y^{y'}d\tl{y}\;\frac{b(\tl{y})}{n(\tl{y})}, 
 \label{def:cU}
\end{align}
and
\begin{align}
 \cT_l^{(tt)}(y,y') &\equiv \frac{H_{l+1}(y,y')}{2}+\frac{n^2(\yh)}{b^2(\yh)}\cS_l(y,y')+n^2(\yh)M^2\cH_{l-1}(y,y'), \nonumber\\
 \cT_l^{(yy)}(y,y') &\equiv H_{l+1}(y,y')-\frac{n^2(\yh)}{b^2(\yh)}\cS_l(y,y')+n^2(\yh)M^2\cH_{l-1}(y,y'). 
\end{align}
Here $\cH_l(y,y')$ and $\cS_l(y,y')$ are defined in Appendix~\ref{apx_expr:boson}, 
and are determined by the background metric.  
Therefore, the divergent parts~$\Lvev{T_{MN}^{\rm b}}^{\rm UV}$ are expressed as
\begin{align}
 \Lvev{T_{tt}^{\rm b}}^{\rm UV}(y,y') &= \frac{\cI^{(tt)}(y,y')}{32\pi^2 n^{3/2}(y)n^{3/2}(y')}, \nonumber\\
 \Lvev{T_{ij}^{\rm b}}^{\rm UV}(y,y') &= -\frac{\asc^2\dlt_{ij}\cI^{(tt)}(y,y')}{32\pi^2 n^{3/2}(y)n^{3/2}(y')}, \nonumber\\
 \Lvev{T_{yy}^{\rm b}}^{\rm UV}(y,y') &= -\frac{b^2(\yh)\cI^{(yy)}(y,y')}{32\pi^2n^2(\yh)n^{3/2}(y)n^{3/2}(y')}, 
\end{align}
where
\begin{align}
 \cI^{(tt)}(y,y') &\equiv \int_{\bar{\rho}}^\infty d\rho\;e^{-\rho\Dlt(y,y')}
 \sbk{\rho^4\cT_{-1}^{(tt)}+\rho^3\cT_0^{(tt)}+\rho^2\cT_1^{(tt)}+\cdots} \nonumber\\
 &= \cW(5,y)\cT_{-1}^{(tt)}+\cW(4,y)\cT_0^{(tt)}+\cW(3,y)\cT_1^{(tt)}+\cW(2,y)\cT_2^{(tt)}+\cdots, 
 \nonumber\\
 \cI^{(yy)}(y,y') &\equiv 
 \cW(5,y)\cT_{-1}^{(yy)}+\cW(4,y)\cT_0^{(yy)}+\cW(3,y)\cT_1^{(yy)}+\cW(2,y)\cT_2^{(yy)}+\cdots, \nonumber\\
 \cW(a,y) &\equiv \frac{\Gm(a,\bar{\rho}\Dlt(y,y'))}{\Dlt^a(y,y')}, 
 \label{def:cW}
\end{align}
and $\Gm(a,x)\equiv \int_x^\infty dw\;e^{-w}w^{a-1}$ is the incomplete gamma function. 
Using the expansion~\eqref{expand:cW}, we obtain
\begin{align}
 \Lvev{T_{tt}^{\rm b}}^{\rm UV} &= \frac{1}{32\pi^2}\sbk{-\frac{12n^2(y)}{b^5(y)}\Lct^5+\cC_{\rm B4}^{(tt)}(y)\Lct^4+\cC_{\rm B3}^{(tt)}(y)\Lct^3
 +\cC_{\rm B2}^{(tt)}(y)\Lct^2+\cdots}, 
 \nonumber\\
 \Lvev{T_{ij}^{\rm b}}^{\rm UV} &= -\frac{\asc^2\dlt_{ij}}{32\pi^2}\sbk{-\frac{12n^2(y)}{b^5(y)}\Lct^5+\cC_{\rm B4}^{(tt)}(y)\Lct^4
 +\cC_{\rm B3}^{(tt)}(y)\Lct^3+\cC_{\rm B2}^{(tt)}(y)\Lct^2+\cdots}, 
 \nonumber\\
 \Lvev{T_{yy}^{\rm b}}^{\rm UV} &= \frac{b^2(y)}{32\pi^2n^2(y)}\sbk{-\frac{48n^2(y)}{b^5(y)}\Lct^5
 +\cC_{\rm B4}^{(yy)}(y)\Lct^4+\cC_{\rm B3}^{(yy)}(y)\Lct^3+\cC_{\rm B2}^{(yy)}(y)\Lct^2+\cdots}, 
 \label{bosonic:T_MN}
\end{align}
where
\begin{align}
 \cC_{\rm B4}^{(tt)}(y) &\equiv -\frac{3n(y)}{b^6(y)}\sbk{b^2(y)\cH_1(y,y)-10n(y)b'(y)+4b(y)n'(y)}, \nonumber\\
 \cC_{\rm B3}^{(tt)}(y) &\equiv -\frac{\cH_2(y,y)-2M^2n^2(y)}{b^3(y)}-\frac{78n^2(y)b^{\prime 2}(y)}{2b^7(y)}
 -\frac{2n(y)}{b^6(y)}\brc{17n'(y)b'(y)+2n(y)b''(y)} \nonumber\\
 &\quad
 +\frac{1}{b^5(y)}\brc{6n(y)b'(y)\cH_1(y,y)-4\brkt{n^{\prime 2}(y)+n(y)n''(y)}} \nonumber\\
 &\quad
 -\frac{1}{2b^4(y)}\brc{3n'(y)\cH_1(y,y)+2n(y)\brkt{\der_y'\cH_1(y,y)+2\der_y\cH_1(y,y)}}, 
 \nonumber\\
 &\vdots 
\end{align}
and
\begin{align}
 \cC_{\rm B4}^{(yy)}(y) &\equiv -\frac{12n(y)}{b^6(y)}\sbk{b^2(y)\cH_1(y,y)-6n(y)b'(y)}, \nonumber\\
 \cC_{\rm B3}^{(yy)}(y) &\equiv -\frac{4\cH_2(y,y)+2M^2n^2(y)}{b^3(y)}-\frac{66n^2(y)b^{\prime 2}(y)}{b^7(y)}
 -\frac{11n(y)}{b^6(y)}\brc{n'(y)b'(y)-2n(y)b''(y)} \nonumber\\
 &\quad
 +\frac{1}{2b^5(y)}\brc{24n(y)b'(y)\cH_1(y,y)+n^{\prime 2}(y)+10n(y)n''(y)} \nonumber\\
 &\quad
 +\frac{2}{b^4(y)}\brc{3n'(y)\cH_1(y,y)-n(y)\brkt{5\der_y'\cH_1(y,y)+\der_y\cH_1(y,y)}}, 
 \nonumber\\
 &\vdots
\end{align}

For example, when $n(y)=e^{-ky}$ and $b(y)=b_{\rm c}$ ($k$, $b_{\rm c}$: constant),\footnote{ 
This should not be understood as a static background but a temporal background configuration during the evolution of the universe. 
Hence the constant~$k$ is not related to the 5D cosmological constant~$\Lmd_{\rm cc}^{\rm (5D)}$, 
in contrast to Ref.~\cite{Randall:1999ee}. 
}
the above expressions become 
\begin{align}
 \Lvev{T^{\rm b}_{tt}}^{\rm UV} &= \frac{e^{-2ky}}{32\pi^2b_{\rm c}^5}\bigg[
 -12\Lct^5+12k\Lct^4-\brkt{8k^2-2b_{\rm c}^2M^2}\Lct^3-k\brkt{-4k^2+2b_{\rm c}^2M^2}\Lct^2 \nonumber\\
 &\hspace{20mm}
 +\frac{6k^4-b_{\rm c}^4M^4}{2}\Lct+\frac{\brkt{b_{\rm c}\bar{\rho}e^{ky}}^5}{10}
 -\frac{\brkt{b_{\rm c}\bar{\rho}e^{ky}}^3}{48}\brkt{-9+4b_{\rm c}^2M^2} \nonumber\\
 &\hspace{20mm}
 -\frac{3b_{\rm c}\bar{\rho}e^{ky}}{256}\brkt{9k^2-4b_{\rm c}^2M^2}\brkt{15k^2+4b_{\rm c}^2M^2} \nonumber\\
 &\hspace{20mm}
 -\frac{k\brkt{122k^4-20k^2b_{\rm c}^2M^2-15b_{\rm c}^4M^4}}{30}+\cdots\bigg], \nonumber\\
 \Lvev{T^{\rm b}_{yy}}^{\rm UV} &= \frac{1}{32\pi^2b_{\rm c}^3}\bigg[
 -48\Lct^5+4\brkt{7k^2+b_{\rm c}^2M^2}\Lct^3-k^2\brkt{9k^2+2b_{\rm c}^2M^2}\Lct
 +\frac{2\brkt{b_{\rm c}\bar{\rho}e^{ky}}^5}{5} \nonumber\\
 &\hspace{20mm}
 -\frac{\brkt{b_{\rm c}\bar{\rho}e^{ky}}^3}{12}\brkt{9k^2-4b_{\rm c}^2M^2}
 +\frac{b_{\rm c}\bar{\rho}e^{ky}}{64}\brkt{9k^2-4b_{\rm c}^2M^2}\brkt{15k^2+4b_{\rm c}^2M^2} \nonumber\\
 &\hspace{20mm}
 +\cdots\bigg]. 
 \label{warped_eg:boson}
\end{align}

The $\bar{\rho}$-dependent terms in the above expressions will be cancelled with 
those of the finite part~$\Lvev{T_{MN}}^{\rm fin}$, which is shown in Sec.~\ref{fin_part}.

\subsection{Spinor sector}
When we keep $\Lct$ finite, \eqref{expr:T_MN:fermion} should be expressed as
\begin{align}
 \Lvev{T^{\rm f}_{tt}} &= -\frac{\asc^3n(\yh)}{16\pi^2n^2(y)n^2(y')}\int_0^\infty d\rho\;\rho^5
 \sbk{\bar{G}_{\chi\chi}(\rho^2,y,y')+\bar{G}_{\lmd\lmd}(\rho^2,y,y')}, \nonumber\\
 \Lvev{T^{\rm f}_{ij}} &= \frac{\asc^5n(\yh)}{16\pi^2n^2(y)n^2(y')}\int_0^\infty d\rho\;\rho^5
 \sbk{\bar{G}_{\chi\chi}(\rho^2,y,y')+\bar{G}_{\lmd\lmd}(\rho^2,y,y')}, \nonumber\\
 \Lvev{T^{\rm f}_{yy}} &= \frac{\asc^3b(\yh)}{8\pi^2n^2(y)n^2(y')}
 \int_0^\infty d\rho\;\rho^3\sbk{\brkt{\der_y-\der_y'}\brkt{\bar{G}_{\chi\lmd}(\rho^2,y,y')-\bar{G}_{\lmd\chi}(\rho^2,y,y')}}. 
\end{align}
From this and \eqref{ap:barG}, the $\Lct$-dependent part of $\Lvev{T^{\rm f}_{tt}}$ is 
\begin{align}
 \Lvev{T^{\rm f}_{tt}}^{\rm UV} &= \frac{\asc^3n(\yh)}{16\pi^2n^2(y)n^2(y')}\int_{\bar{\rho}}^\infty d\rho\;\rho^5\bigg[
 \frac{e^{-\rho\Dlt(y,y')}}{2\asc^3\rho}\brkt{1+\frac{2\cK_1^{\chi\chi}}{\rho}+\frac{2\cK_2^{\chi\chi}}{\rho^2}+\cdots} \nonumber\\
 &\hspace{55mm}
 +\frac{e^{-\rho\Dlt(y,y')}}{2\asc^3\rho}\brkt{1+\frac{2\cK_1^{\lmd\lmd}}{\rho}+\frac{2\cK_2^{\lmd\lmd}}{\rho^2}+\cdots}\bigg] \nonumber\\
 &= \frac{n(\yh)}{16\pi^2n^2(y)n^2(y')}\bigg[\cW(5,y)
 +\cW(4,y)\cV_1^{(tt)}(y,y')+\cW(3,y)\cV_2^{(tt)}(y,y')
 \nonumber\\
 &\hspace{35mm}
 +\cW(2,y)\cV_3^{(tt)}(y,y')+\cdots\bigg], \nonumber\\
 \Lvev{T_{yy}^{\rm f}}^{\rm UV} &= -\frac{\asc^3b(\yh)}{8\pi^2n^2(y)n^2(y')}\int_{\bar{\rho}}^\infty d\rho\;\rho^3
 \left[-\frac{e^{-\rho\Dlt(y,y')}}{\asc^3}\brkt{\rho\Xi+\Xi\cV_1^{(yy)}+\frac{\cV_2^{(yy)}}{\rho}+\frac{\cV_3^{(yy)}}{\rho^2}+\cdots}
 \right] \nonumber\\
 &= \frac{b(\yh)}{8\pi^2n^2(y)n^2(y')}\bigg[\cW(5,y)\Xi(y,y')+\cW(4,y)\brkt{\Xi\cV_1^{(yy)}}(y,y') \nonumber\\
 &\hspace{35mm}
 +\cW(3,y)\brkt{\Xi\cV_2^{(yy)}}(y,y')+\cW(2,y)\brkt{\Xi\cV_3^{(yy)}}(y,y')+\cdots \bigg], 
\end{align}
where $y'=y+1/\Lct$, and 
\begin{align}
 \cV_m^{(tt)}(y,y') &\equiv \cK_m^{\chi\chi}(y,y')+\cK_m^{\lmd\lmd}(y,y'), \;\;\;\;\; \brkt{m=1,2,3,\cdots} \nonumber\\
 \Xi(y,y') &\equiv -\der_y\Dlt(y,y')+\der_y'\Dlt(y,y') = \frac{b(y)}{n(y)}+\frac{b(y')}{n(y')}, \nonumber\\
 \cV_1^{(yy)}(y,y') &\equiv \Xi(y,y')\sbk{\cK_1^{\chi\lmd}(y,y')+\cK_1^{\lmd\chi}(y,y')}, \nonumber\\
 \cV_m^{(yy)}(y,y') &\equiv \Xi(y,y')\sbk{\cK_m^{\chi\lmd}(y,y')+\cK_m^{\lmd\chi}(y,y')} \nonumber\\
 &\quad
 +\brkt{\der_y-\der_y'}\sbk{\cK_{m-1}^{\chi\lmd}(y,y')+\cK_{m-1}^{\lmd\chi}(y,y')}. \;\;\;\;\; \brkt{m=2,3,\cdots}
\end{align}
Using the expansion~\eqref{expand:cW}, we obtain
\begin{align}
 \Lvev{T_{tt}^{\rm f}}^{\rm UV} &= \frac{1}{8\pi^2}\sbk{
 \frac{12n^2(y)}{b^5(y)}\Lct^5+\cC_{\rm F4}^{(tt)}(y)\Lct^4+\cC_{\rm F3}^{(tt)}(y)\Lct^3+\cC_{\rm F2}^{(tt)}(y)\Lct^2+\cdots}, \nonumber\\
 \Lvev{T_{ij}^{\rm f}}^{\rm UV} &= -\frac{\asc^2}{8\pi^2}\sbk{
 \frac{12n^2(y)}{b^5(y)}\Lct^5+\cC_{\rm F4}^{(tt)}(y)\Lct^4+\cC_{\rm F3}^{(tt)}(y)\Lct^3+\cC_{\rm F2}^{(tt)}(y)\Lct^2+\cdots}, \nonumber\\
 \Lvev{T_{yy}^{\rm f}}^{\rm UV} &= \frac{b^2(y)}{8\pi^2n^2(y)}\sbk{\frac{48n^2(y)}{b^5(y)}\Lct^5
 +\cC_{\rm F4}^{(yy)}(y)\Lct^4+\cC_{\rm F3}^{(yy)}(y)\Lct^3+\cC_{\rm F2}^{(yy)}(y)\Lct^2+\cdots}, 
 \label{fermionic:T_MN}
\end{align}
where
\begin{align}
 \cC_{\rm F4}^{(tt)}(y) &\equiv \frac{3}{b^6(y)}\sbk{
 n(y)b^2(y)\cV_1^{(tt)}(y,y)-10n(y)b'(y)+4b(y)n'(y)}, \nonumber\\
 \cC_{\rm F3}^{(tt)}(y) &\equiv \frac{\cV_2^{(tt)}(y,y)}{b^2(y)}+\frac{45n^2(y)b^{\prime 2}(y)}{b^7(y)}
 -\frac{5n(y)}{b^6(y)}\brc{5n'(y)b'(y)+2n(y)b''(y)} \nonumber\\
 &\quad
 -\frac{1}{2b^5(y)}\brc{12n(y)b'(y)\cV_1^{(tt)}(y,y)-8n^{\prime 2}(y)+n(y)n''(y)} \nonumber\\
 &\quad
 +\frac{3}{2b^4(y)}\brc{n'(y)\cV_1^{(tt)}(y,y)+2n(y)\der_y'\cV_1^{(tt)}(y,y)}, 
 \nonumber\\
 &\vdots, 
\end{align}
and
\begin{align}
 \cC_{\rm F4}^{(yy)}(y) &\equiv \frac{6}{b^6(y)}\sbk{
 n^2(y)b(y)\cV_1^{(yy)}(y,y)-12n^2(y)b'(y)}, \nonumber\\
 \cC_{\rm F3}^{(yy)}(y) &\equiv \frac{2n(y)}{b^7(y)}\bigg[\cV_2^{(yy)}(y,y)
 +\frac{3b^2(y)}{2}\brc{n'(y)\cV_1^{(yy)}(y,y)+2n(y)\der_y'\cV_1^{(yy)}(y,y)} \nonumber\\
 &\hspace{20mm}
 +36n(y)b^{\prime 2}(y)+b(y)\brkt{4n'(y)b'(y)-11n(y)b''(y)}\bigg],  
 \nonumber\\
 &\vdots
\end{align}

For example, when $n(y)=e^{-ky}$ and $b(y)=b_{\rm c}$, the above expressions become
\begin{align}
 \Lvev{T^{\rm f}_{tt}}^{\rm UV} &= \frac{e^{-2ky}}{8\pi^2b_{\rm c}^5}\bigg[
 12\Lct^5-12k\Lct^4+\frac{7k^2-4b_{\rm c}^2M^2}{2}\Lct^3
 +\frac{k\brkt{k^2+4b_{\rm c}^2M^2}}{2}\Lct^2 \nonumber\\
 &\hspace{17mm}
 -\frac{15k^4+40k^2b_{\rm c}^2M^2-16b_{\rm c}^4M^4}{32}\Lct \nonumber\\
 &\hspace{17mm}
 +\frac{1}{480}\bigg\{k\brkt{17k^4+90kb_{\rm c}^3\bar{\rho}e^{ky}M^2+280k^2b_{\rm c}^2M^2-240b_{\rm c}^4M^4} \nonumber\\
 &\hspace{30mm}
 -b_{\rm c}^5\brkt{48\bar{\rho}^5e^{5ky}-40\bar{\rho}^3e^{3ky}M^2+90\bar{\rho}e^{ky}M^4}\bigg\}+\cdots\bigg], \nonumber\\
 \Lvev{T^{\rm f}_{yy}}^{\rm UV} &= \frac{1}{8\pi^2b_{\rm c}^3}\bigg[
 48\Lct^5-4\brkt{k^2+b_{\rm c}^2M^2}\Lct^3 \nonumber\\
 &\hspace{17mm}
 -\frac{b_{\rm c}\bar{\rho}e^{ky}}{60}\brc{24\brkt{b_{\rm c}\bar{\rho}e^{ky}}^4+20\brkt{b_{\rm c}\bar{\rho}e^{ky}}^2b_{\rm c}^2M^2
 +15b_{\rm c}^2M^2\brkt{k^2-b_{\rm c}^2M^2}}+\cdots\bigg]. 
 \label{warped_eg:fermion}
\end{align}

\subsection{Total energy-momentum tensor} \label{total_EMT}
Here we consider a situation that the conditions in \eqref{Pauli_cond} are satisfied. 
Specifically, we have $N_{\rm B}$ 5D scalar fields~$\Phi_i$ with masses~$M_{{\rm B}i}$ ($i=1,2,\cdots,N_{\rm B}$) 
and $N_{\rm F}$ 5D spinor fields~$\Psi_i$ with masses~$M_{{\rm F}i}$ ($i=1,2,\cdots,N_{\rm F}$), that satisfy 
\begin{align}
 N_{\rm B}-4N_{\rm F} &= 0, \nonumber\\
 \sum_{i=1}^{N_{\rm B}}M_{{\rm B}i}^2-4\sum_{i=1}^{N_{\rm F}}M_{{\rm F}i}^2 &=0, \nonumber\\
\sum_{i=1}^{N_{\rm B}}M_{{\rm B}i}^4-4\sum_{i=1}^{N_{\rm F}}M_{{\rm F}i}^4 &= 0. 
\label{eg:Pauli_cond}
\end{align}
We do not assume any hierarchies among $M_{{\rm B}i}$ and $M_{{\rm F}i}$, unlike we did in Sec.~\ref{review}. 
From \eqref{bosonic:T_MN} and \eqref{fermionic:T_MN}, the $\Lct$-dependent part of $\Lvev{T_{MN}}$ has the following form. 
\begin{align}
 \Lvev{T_{tt}}^{\rm UV} &= \frac{1}{32\pi^2}\left[
 \frac{12n^2(y)}{b^5(y)}\brkt{-N_{\rm B}+4N_{\rm F}}\Lct^5
 +C_4^{(tt)}(y)\Lct^4+C_3^{(tt)}(y)\Lct^3+C_2^{(tt)}(y)\Lct^2+\cdots\right], \nonumber\\
 \Lvev{T_{yy}}^{\rm UV} &= \frac{b^2(y)}{32\pi^2n^2(y)}\left[\frac{48n^2(y)}{b^5(y)}\brkt{-N_{\rm B}+4N_{\rm F}}\Lct^5
 +C_4^{(yy)}(y)\Lct^4+C_3^{(yy)}(y)\Lct^3+\cdots\right], 
 \label{expr:totalT_MN}
\end{align}
where
\begin{align}
 C^{(tt)}_k(y) &\equiv \sum_{i=1}^{N_{\rm B}}\cC_{{\rm B}k}^{(tt)}(y;M_{{\rm B}i})+4\sum_{i=1}^{N_{\rm F}}\cC_{{\rm F}k}^{(tt)}(y;M_{{\rm F}i}), \nonumber\\
 C^{(yy)}_k(y) &\equiv \sum_{i=1}^{N_{\rm B}}\cC_{{\rm B}k}^{(yy)}(y;M_{{\rm B}i})+4\sum_{i=1}^{N_{\rm F}}\cC_{{\rm F}k}^{(yy)}(y;M_{{\rm F}i}). \;\;\;\;\;
 \brkt{k=4,3,2,1,\cdots}
\end{align}
The coefficient  functions~$\brc{\cC_{{\rm B}k}^{(tt)},\cC_{{\rm B}k}^{(yy)}}$ and $\brc{\cC_{{\rm F}k}^{(tt)},\cC_{{\rm F}k}^{(yy)}}$  
depend on the masses~$M_{{\rm B}i}$ and $M_{{\rm F}i}$, respectively. 
Since the non-vanishing components of $\Lvev{T_{ij}}^{\rm UV}$ ($i,j=1,2,3$) are proportional to $\Lvev{T_{tt}}^{\rm UV}$, 
we do not show them here. 

Note that the leading $\cO(\Lct^5)$ terms are always cancelled due to the first condition in \eqref{eg:Pauli_cond}. 
Examining $\cC_{{\rm B,F}4}^{(tt)}$ and $\cC_{{\rm B,F}4}^{(yy)}$ for various background geometries, 
we can see that they also vanish under the first condition in \eqref{eg:Pauli_cond}. 
The explicit forms of $\cC_{{\rm B,F}k}^{(tt)}$ and $\cC_{{\rm B,F}k}^{(yy)}$ ($k=4,3,2,\cdots$) 
in terms of the metric components~$n(y)$ and $b(y)$ are quite complicated and lengthy. 
Hence we do not show their explicit expressions here. 
Instead, we will provide the expressions in some specific cases below.

\begin{description}
\item[Case of {\boldmath $n(y)=1+cy^2$} and {\boldmath $b(y)=b_{\rm c}$} ($\bdm{c}$, $\bdm{b_{\rm c}}$ : constant)]
\begin{align}
 C_4^{(tt)}(y) &= \frac{24cy\brkt{1+cy^2}}{b_{\rm c}^5}\brkt{-N_{\rm B}+4N_{\rm F}}, \nonumber\\
 C_3^{(tt)}(y) &= -\frac{4c\brkt{2N_{\rm B}+N_{\rm F}}+4c^2y^2\brkt{6N_{\rm B}-15N_{\rm F}}}{b_{\rm c}^5}
 +\frac{2\brkt{1+cy^2}^2}{b_{\rm c}^3}\brkt{\sum_{i=1}^{N_{\rm F}}M_{{\rm B}i}^2-4\sum_{i=1}^{N_{\rm F}}M_{{\rm F}i}^2}, 
 \nonumber\\
 C_2^{(tt)}(y) &= -\frac{12c^2y\brkt{N_{\rm B}-N_{\rm F}}}{b_{\rm c}^5}
 +\frac{4cy\brkt{1+cy^2}}{b_{\rm c}^3}\brkt{\sum_{i=1}^{N_{\rm F}}M_{{\rm B}i}^2-4\sum_{i=1}^{N_{\rm F}}M_{{\rm F}i}^2}, \nonumber\\
 &\vdots
\end{align}
and
\begin{align}
 C_4^{(yy)}(y) &= 0, \nonumber\\
 C_3^{(yy)}(y) &= \frac{4c(7N_{\rm B}-40N_{\rm F})+12c^2y^2(7N_{\rm B}+8N_{\rm F})}{b_{\rm c}^5} \nonumber\\
 &\quad
 +\frac{4\brkt{1+cy^2}^2}{b_{\rm c}^3}\brkt{\sum_{i=1}^{N_{\rm F}}M_{{\rm B}i}^2-4\sum_{i=1}^{N_{\rm F}}M_{{\rm F}i}^2}, \nonumber\\
 C_2^{(yy)}(y) &= -\frac{4c^2y}{b_{\rm c}^5\brkt{1+cy^2}}\sbk{-7N_{\rm B}-104N_{\rm F}+3cy^2\brkt{7N_{\rm B}+8N_{\rm F}}}, \nonumber\\
 &\vdots
\end{align}

\item[Case of {\boldmath $n(y)=e^{-ky}$} and {\boldmath $b(y)=b_{\rm c}$} ($\bdm{k}$, $\bdm{b_{\rm c}}$ : constant)] \mbox{}\\
\begin{align}
 C_4^{(tt)}(y) &= \frac{12e^{-2ky}k\brkt{N_{\rm B}-4N_{\rm F}}}{b_{\rm c}^5}, \nonumber\\
 C_3^{(tt)}(y) &= -\frac{2e^{-2ky}k^2\brkt{4N_{\rm B}-7N_{\rm F}}}{b_{\rm c}^5}
 +\frac{2e^{-2ky}}{b_{\rm c}^3}\brkt{\sum_{i=1}^{N_{\rm F}}M_{{\rm B}i}^2-4\sum_{i=1}^{N_{\rm F}}M_{{\rm F}i}^2}, \nonumber\\
 C_2^{(tt)}(y) &= \frac{2e^{-2ky}k^3\brkt{2N_{\rm B}+N_{\rm F}}}{b_{\rm c}^5}
 -\frac{2e^{-2ky}k}{b_{\rm c}^3}\brkt{\sum_{i=1}^{N_{\rm F}}M_{{\rm B}i}^2-4\sum_{i=1}^{N_{\rm F}}M_{{\rm F}i}^2}, \nonumber\\
 &\vdots 
 \label{warped_case:tt}
\end{align}
and
\begin{align}
 C_4^{(yy)}(y) &= 0, \nonumber\\
 C_3^{(yy)}(y) &= \frac{4e^{-2ky}k^2\brkt{7N_{\rm B}-4N_{\rm F}}}{b_{\rm c}^5}
 +\frac{4e^{-2ky}}{b_{\rm c}^3}\brkt{\sum_{i=1}^{N_{\rm F}}M_{{\rm B}i}^2-4\sum_{i=1}^{N_{\rm F}}M_{{\rm F}i}^2}, 
 \nonumber\\
 C_2^{(yy)}(y) &= 0 , \nonumber\\
 &\vdots
 \label{warped_case:yy}
\end{align}
We can check this result from \eqref{warped_eg:boson} and \eqref{warped_eg:fermion}. 
\ignore{
\begin{align}
 \Lvev{T_{tt}}^{\rm UV} &= \frac{e^{-2ky}}{32\pi^2b_{\rm c}^5}\bigg[
 2\brkt{4N_{\rm B}-7N_{\rm F}}k^2\Lct^3
 -2\brkt{2N_{\rm B}+N_{\rm F}}k^3\Lct^2 \nonumber\\
 &\hspace{20mm}
 +\brc{\brkt{-3N_{\rm B}+\frac{15}{8}N_{\rm F}}k^4+5k^2b_{\rm c}^2\sum_{i=1}^{N_{\rm F}}M_{{\rm F}i}^2}\Lct+\cdots\bigg], 
 \nonumber\\
 \Lvev{T_{yy}}^{\rm UV} &= \frac{1}{32\pi^2b_{\rm c}^3}\sbk{
 4\brkt{-7N_{\rm B}+4N_{\rm F}}k^2\Lct^3
 +\brc{9N_{\rm B}k^4+2k^2b_{\rm c}^2\sum_{i=1}^{N_{\rm B}}M_{{\rm B}i}^2}\Lct+\cdots}. 
\end{align}
We have used the conditions in \eqref{eg:Pauli_cond}. 
Notice that all terms wth positive powers in $\Lct$ have a common $y$-dependence. 
In contrast, the $\Lct$-independent terms have non-trivial $y$-dependences. 
}
\end{description}

If the conditions in \eqref{eg:Pauli_cond} are satisfied, 
the coefficients~$C_k^{(tt)}(y)$ and $C_k^{(yy)}(y)$ ($k=4,3,2,1$) vanish 
in the flat limit~$c\to 0$ (or $k\to 0$), as expected. 
Namely, the result is 
\begin{align}
 \Lvev{T_{tt}}^{\rm UV} &= \cO\brkt{\frac{1}{\Lct}}, \;\;\;\;\;
 \Lvev{T_{yy}}^{\rm UV} = \cO\brkt{\frac{1}{\Lct}}. 
 \label{T_MN^UV}
\end{align}

\subsection{Finite part} \label{fin_part}
Here we collect the expressions of the finite part of $\Lvev{T_{tt}}^{\rm fin}$. 
They are expressed as
\begin{align}
 \Lvev{T^{\rm b}_{tt}}^{\rm fin} &= \frac{b(L)}{16\pi^2n^4(L)} \nonumber\\
 &\quad
 \times\int_0^{\bar{\rho}}d\rho\;\rho^3\sbk{\brkt{\frac{\rho^2}{2}+n^2(y)M^2}\frac{C_0(y;\rho^2)C_L(y;\rho^2)}{C_0'(L;\rho^2)}
 +\frac{n^2(y)}{b^2(y)}\frac{C_0'(y;\rho^2)C_L'(y;\rho^2)}{C_0'(L;\rho^2)}}, \nonumber\\
 \Lvev{T_{yy}^{\rm b}}^{\rm fin} &= -\frac{b(L)}{16\pi^2n^4(L)}\int_0^{\bar{\rho}} d\rho\;\rho^3
 \sbk{\brkt{\rho^2+n^2(y)M^2}\frac{b^2(y)C_0(y;\rho^2)C_L(y;\rho^2)}{n^2(y)C_0'(L;\rho^2)}
 -\frac{C_0'(y;\rho^2)C_L'(y;\rho^2)}{C_0'(L;\rho^2)}}, 
 \nonumber\\
 \Lvev{T^{\rm f}_{tt}}^{\rm fin} &= \frac{n(L)}{16\pi^2n^3(y)b(L)}\int_0^{\bar{\rho}}d\rho\;\rho^5
 \sbk{\frac{H_{0\chi}(y;\rho^2)H_{L\chi}(y;\rho^2)}{S_L^{(-)}(0;\rho^2)}-\frac{H_{0\lmd}(y;\rho^2)H_{L\lmd}(y;\rho^2)}{\rho^2S_L^{(-)}(0;\rho^2)}}, \nonumber\\
 \Lvev{T_{yy}^{\rm f}}^{\rm fin} &= -\frac{n(L)}{8\pi^2n^4(y)b(L)}\int_0^{\bar{\rho}}d\rho\;\rho^3\left[
 \frac{H_{0\chi}'(y;\rho^2)H_{L\lmd}(y;\rho^2)-H_{0\chi}(y;\rho^2)H_{L\lmd}'(y;\rho^2)}{S_L^{(-)}(0;\rho^2)} \right.\nonumber\\
 &\hspace{50mm}\left. 
 -\frac{H_{0\lmd}'(y;\rho^2)H_{L\chi}(y;\rho^2)-H_{0\lmd}(y;\rho^2)H_{L\chi}'(y;\rho^2)}{S_L^{(-)}(0;\rho^2)}\right], 
\end{align}
where $H_{u\chi}(y;\rho^2)$ and $H_{u\lmd}(y;\rho^2)$ ($u=0,L$) are defined in \eqref{def:H0L}. 

Let us consider the flat case, $n(y)=1$ and $b(y)=b_{\rm c}$, as an example. 
Then this becomes 
\begin{align}
 \Lvev{T_{tt}^{\rm b}}^{\rm fin} &= \int_{M_{\rm B}}^{\sqrt{\bar{\rho}^2+M_{\rm B}^2}}\frac{d\hat{\rho}}{16\pi^2}\;
 \brkt{\hat{\rho}^2-M_{\rm B}^2} \nonumber\\
 &\hspace{35mm}\times
 \sbk{\frac{3\hat{\rho}^2+M_{\rm B}^2}{4}\frac{\cosh\brkt{b_{\rm c}\hat{\rho}(2y-L)}}{\sinh(b_{\rm c}\hat{\rho}L)}
 -\frac{\hat{\rho}^2-M_{\rm B}^2}{4}\coth\brkt{b_{\rm c}\hat{\rho}L}}, \nonumber\\
 \Lvev{T_{yy}^{\rm b}}^{\rm fin} &= -\frac{b_{\rm c}^2}{16\pi^2}\int_{M_{\rm B}}^{\sqrt{\bar{\rho}^2+M_{\rm B}^2}}d\hat{\rho}\;
 \hat{\rho}^2\brkt{\hat{\rho}^2-M_{\rm B}^2}\coth\brkt{b_{\rm c}\hat{\rho}L}, \nonumber\\
 \Lvev{T_{tt}^{\rm f}}^{\rm fin} &= \int_{M_{\rm F}}^{\sqrt{\bar{\rho}^2+M_{\rm F}^2}}\frac{d\hat{\rho}}{16\pi^2}\;
 \left[\brkt{\hat{\rho}^2-M_{\rm F}^2}\frac{M_{\rm F}^2\cosh\brkt{b_{\rm c}\hat{\rho}(2y-L)}
 -M_{\rm F}\hat{\rho}\sinh\brkt{b_{\rm c}\hat{\rho}(2y-L)}}{\sinh(b_{\rm c}\hat{\rho}L)} \right. \nonumber\\
 &\hspace{37mm}\left.
 +\brkt{\hat{\rho}^2-M_{\rm F}^2}^2\coth\brkt{b_{\rm c}\hat{\rho}L}\right], 
 \nonumber\\
 \Lvev{T_{yy}^{\rm f}}^{\rm fin} &= \frac{b_{\rm c}^2}{4\pi^2}\int_{M_{\rm F}}^{\sqrt{\bar{\rho}^2+M_{\rm F}^2}}d\hat{\rho}\;
 \hat{\rho}^2\brkt{\hat{\rho}^2-M_{\rm F}^2}\coth\brkt{b_{\rm c}\hat{\rho}L}, 
 \label{flat:T_MN}
\end{align}
where $\hat{\rho}\equiv \sqrt{\rho^2+M_{\rm B,F}^2}$. 
Note that the $y$-dependent parts in $\Lvev{T_{tt}^{\rm b}}^{\rm fin}$ and $\Lvev{T_{tt}^{\rm f}}^{\rm fin}$ stem from 
the existence of the boundaries at $y=0,L$. 
In fact, in the case of the $S^1$ compactification, they are absent. 
Except in the vicinity of the boundaries, their contributions are exponentially suppressed and negligible. 
Then, when $M_{\rm B}b_{\rm c}L,\; M_{\rm F}b_{\rm c}L\gg 1$, 
\eqref{flat:T_MN} are approximated as
\begin{align}
 \Lvev{T_{tt}^{\rm b}}^{\rm fin} &\simeq -\Lvev{T_{tt}}_0(M_{\rm B}), \;\;\;\;\;
 \Lvev{T_{tt}^{\rm f}}^{\rm fin} \simeq 4\Lvev{T_{tt}}_0(M_{\rm F}), \nonumber\\
 \Lvev{T_{yy}^{\rm b}}^{\rm fin} &\simeq -\Lvev{T_{yy}}_0(M_{\rm B}), \;\;\;\;\;
 \Lvev{T_{yy}^{\rm f}}^{\rm fin} \simeq 4\Lvev{T_{yy}}_0(M_{\rm F}), 
\end{align}
where
\begin{align}
 \Lvev{T_{tt}}_0(M) &\equiv \frac{1}{64\pi^2}\int_M^{\sqrt{\bar{\rho}^2+M^2}}d\hat{\rho}\;\brkt{\hat{\rho}^2-M^2}^2 \nonumber\\
 &\simeq \frac{1}{64\pi^2}\brkt{\frac{\bar{\rho}^5}{5}-\frac{M^2\bar{\rho}^3}{6}+\frac{3M^4\bar{\rho}}{8}-\frac{8M^5}{15}
 +\cO\brkt{\frac{1}{\bar{\rho}}}}, \nonumber\\
 \Lvev{T_{yy}}_0(M) &\equiv \frac{b_{\rm c}^2}{16\pi^2}\int_M^{\sqrt{\hat{\rho}^2+M^2}}d\hat{\rho}\;\hat{\rho}^2\brkt{\bar{\rho}^2-M^2} \nonumber\\
 &= \frac{b_{\rm c}^2}{32\pi^2}\brkt{\frac{2\bar{\rho}^5}{5}+\frac{M^2\bar{\rho}^3}{3}-\frac{M^4\bar{\rho}}{4}+\frac{4M^5}{15}+\cO\brkt{\frac{1}{\bar{\rho}}}}. 
\end{align}
We can easily check that the positive-power terms in $\bar{\rho}$ in $\Lvev{T_{tt}}_0$ and $\Lvev{T_{yy}}_0$ are cancelled 
with those in $\Lvev{T_{tt}}^{\rm UV}$ and $\Lvev{T_{yy}}^{\rm UV}$, which are obtained by taking the limit~$k\to 0$ 
in \eqref{warped_eg:boson} and \eqref{warped_eg:fermion}. 
Therefore, in the system considered in Sec.~\ref{total_EMT}, the total contributions are calculated as
\begin{align}
 \Lvev{T_{tt}}^{\rm fin} &= \sum_{i=1}^{N_{\rm B}}\Lvev{T_{tt}^{\rm b}}^{\rm fin}(M_{{\rm B}i})+\sum_{i=1}^{N_{\rm F}}\Lvev{T_{tt}^{\rm f}}^{\rm fin}(M_{{\rm F}i}) 
 \nonumber\\
 &\simeq -\sum_{i=1}^{N_{\rm B}}\Lvev{T_{tt}}_0(M_{{\rm B}i})+4\sum_{i=1}^{N_{\rm F}}\Lvev{T_{tt}}_0(M_{{\rm F}i}) 
 \nonumber\\
 &= \frac{\bar{\rho}^5(-N_{\rm B}+4N_{\rm F})}{320\pi^2}
 +\frac{\bar{\rho}^3}{384\pi^2}\brkt{\sum_{i=1}^{N_{\rm B}}M_{{\rm B}i}^2-4\sum_{i=1}^{N_{\rm F}}M_{{\rm F}i}^2}
 +\frac{3\bar{\rho}}{512\pi^2}\brkt{\sum_{i=1}^{N_{\rm B}}M_{{\rm B}i}^4-4\sum_{i=1}^{N_{\rm F}}M_{{\rm F}i}^4} \nonumber\\
 &\quad
 +\frac{1}{120\pi^2}\brkt{\sum_{i=1}^{N_{\rm B}}M_{{\rm B}i}^5-4\sum_{i=1}^{N_{\rm F}}M_{{\rm F}i}^5}+\cO\brkt{\frac{1}{\bar{\rho}}} 
 \nonumber\\
 &= \frac{1}{120\pi^2}\brkt{\sum_{i=1}^{N_{\rm B}}M_{{\rm B}i}^5-4\sum_{i=1}^{N_{\rm F}}M_{{\rm F}i}^5}+\cO\brkt{\frac{1}{\bar{\rho}}}, 
\end{align}
At the last step, we have used the conditions in \eqref{eg:Pauli_cond}. 
Similarly, we have
\begin{align}
 \Lvev{T_{yy}}^{\rm fin} &\simeq \frac{b_{\rm c}^2}{120\pi^2}\brkt{-\sum_{i=1}^{N_{\rm B}}M_{{\rm B}i}^5
 +4\sum_{i=1}^{N_{\rm F}}M_{{\rm F}i}^5}+\brkt{\frac{1}{\bar{\rho}}}. 
\end{align}
Combining these results with \eqref{T_MN^UV}, we obtain
\begin{align}
 \Lvev{T_{MN}} &\simeq \frac{1}{120\pi^2}\brkt{-\sum_{i=1}^{N_{\rm B}}M_{{\rm B}i}^5+4\sum_{i=1}^{N_{\rm F}}M_{{\rm F}i}^5}\times
 \begin{pmatrix} -1 & & \\ & \asc^2\id_3 & \\ & & b_{\rm c}^2 \end{pmatrix}+\cO\brkt{\frac{1}{\bar{\rho}}}. 
 \label{finite_result}
\end{align}
Namely, $\Lvev{T_{MN}}\propto g_{MN}$, and thus this contributes to the renormalization of the 5D cosmological constant. 
This is the result in the flat case. 

Before ending this section, we will see the relation to the conventional result~\eqref{cv_result:massive} or \eqref{cv_result:massless}. 
In fact, it corresponds to the following redefinition of  (the VEV of) the energy-momentum tensor. 
\begin{align}
 \Lvev{\hat{T}_{MN}}^{\rm fin} &\equiv \Lvev{T_{MN}}_{\rm const}^{\rm fin}-\lim_{L\to\infty}\Lvev{T_{MN}}_{\rm const}^{\rm fin} \nonumber\\
 &= \Lvev{T_{MN}}_{\rm const}^{\rm fin}-\brkt{-\sum_{i=1}^{N_{\rm B}}\Lvev{T_{MN}}_0(M_{{\rm B}i})
 +4\sum_{i=1}^{N_{\rm F}}\Lvev{T_{MN}}_0(M_{{\rm F}i})}. 
 \label{modifiedT_MN}
\end{align}
where $\Lvev{T_{MN}}_{\rm const}^{\rm fin}$ is the constant part of $\Lvev{T_{MN}}^{\rm fin}$. 
Then, we have
\begin{align}
 \Lvev{\hat{T}_{tt}}^{\rm fin} &= -\sum_{i=1}^{N_{\rm B}}\cF_{tt}(M_{{\rm B}i})
 +4\sum_{i=1}^{N_{\rm F}}\cF_{tt}(M_{{\rm F}i}), \nonumber\\
 \Lvev{\hat{T}_{yy}}^{\rm fin} &= -\sum_{i=1}^{N_{\rm B}}\cF_{yy}(M_{{\rm B}i})
 +4\sum_{i=1}^{N_{\rm F}}\cF_{yy}(M_{{\rm F}i}),
\end{align}
where
\begin{align}
 \cF_{tt}(M) &\equiv \frac{1}{64\pi^2}\int_{M}^{\sqrt{\bar{\rho}^2+M^2}}d\hat{\rho}\;
 \brkt{\hat{\rho}^2-M^2}^2\sbk{\coth\brkt{b_{\rm c}\hat{\rho}L}-1} \nonumber\\
 &= \frac{M^5}{4\pi^2}\sbk{\frac{1}{\alp^3}{\rm Li}_3(e^{-\alp})+\frac{3}{\alp^4}{\rm Li}_4(e^{-\alp})+\frac{3}{\alp^5}(e^{-\alp})}
 +\cO\brkt{e^{-2b_{\rm c}\bar{\rho}L}}, \nonumber\\
 \cF_{yy}(M) &\equiv \frac{b_{\rm c}^2}{16\pi^2}\int_M^{\sqrt{\bar{\rho}^2+M^2}}d\hat{\rho}\;
 \hat{\rho}^2\brkt{\hat{\rho}^2-M^2}\sbk{\coth\brkt{b_{\rm c}\hat{\rho}L}-1} \nonumber\\
 &= \frac{b_{\rm c}^2M^5}{4\pi^2}\sbk{\frac{1}{\alp^2}{\rm Li}\brkt{e^{-\alp}}+\frac{5}{\alp^3}{\rm Li}_3\brkt{e^{-\alp}}
 +\frac{12}{\alp^4}{\rm Li}_4\brkt{e^{-\alp}}+\frac{12}{\alp^5}{\rm Li}_5\brkt{e^{-\alp}}}
 +\cO\brkt{e^{-2b_{\rm c}\bar{\rho}L}}, 
\end{align}
with $\alp\equiv 2b_{\rm c}LM$. 
In the massless limit~$M\to 0$, this becomes
\begin{align}
 \cF_{tt}(0) &= \frac{3\zeta(5)}{128\pi^2(b_{\rm c}L)^5}+\cO\brkt{e^{-2b_{\rm c}\bar{\rho}L}}, \;\;\;\;\;
 \cF_{yy}(0) = \frac{3\zeta(5)b_{\rm c}^2}{32\pi^2(b_{\rm c}L)^5}+\cO\brkt{e^{-2b_{\rm c}\bar{\rho}L}}. 
 \label{cv_result:flat}
\end{align}
If we rewrite the coordinate distance between the boundaries as $L=\pi R$, 
the above results are consistent with
the conventional results~\eqref{cv_result:massive} and \eqref{cv_result:massless} in 4D effective theory. 
However, the origin of the second term in the modification~\eqref{modifiedT_MN} is still unclear.

\section{Summary and discussions} \label{summary}
We have calculated the cutoff-dependence of (the VEV of) the 5D energy-momentum tensor~$\Lvev{T_{MN}}$ 
in terms of the metric components~$n(y)$ and $b(y)$ in \eqref{metric_ansatz}. 
We have kept the cutoff energy scale~$\Lct$ finite, and investigate the possibility that 
the $\Lct$-dependence is cancelled among contributions of different fields. 
It is known that this actually occurs if the conditions in \eqref{eg:Pauli_cond} are satisfied in the 4D effective theory. 
The purpose of this paper is to check whether this cancellation occurs in the 5D setup. 
Note that the first condition in \eqref{eg:Pauli_cond} is easily satisfied in a case that the supersymmetry exists at least at high energies 
even if it is broken at low energies. 
Hence we focus on such a case in the following. 
The results obtained in Sec.~\ref{total_EMT} show that 
\begin{enumerate}
 \item The $\cO(\Lct^5)$- and the $\cO(\Lct^4)$-contributions are always cancelled.  
 
 \item In the flat spacetime ($n(y)=1$ and $b(y)=b_{\rm c}$), all terms with positive powers in $\Lct$ are cancelled 
 if the mass relations in \eqref{eg:Pauli_cond} are satisfied. 
 
 \item In the slice of AdS space ($n(y)=e^{-ky}$ and $b(y)=b_{\rm c}$), $\cO(\Lct^3)$-contributions are not cancelled and survive. 
 However, we should note that all the coefficients for the positive power of $\Lct$ have a common $y$-dependence 
 (i.e., $C_k^{(tt)}(y), \, C_k^{(yy)}(y) \propto e^{-2ky}$, where $k=4,3,2,1$). 
 Thus there is a possibility that they are cancelled if the mass relations in \eqref{eg:Pauli_cond} are modified appropriately. 
 \label{comment3}
 
 \item In other background geometries, $\cO(\Lct^3)$-contributions always survive. 
 In contrast to the above cases, the $y$-dependence of the bosonic and the fermionic contributions are different. 
 Thus there is no chance to cancel all the coefficients no matter how the mass relations are modified. 
 \label{comment4}
\end{enumerate}
Although only the case of $\brc{n(y)=1+cy^2,\,b(y)=b_{\rm c}}$ is shown as an example of non-AdS space in Sec.~\ref{total_EMT}, 
we have checked that the properties mentioned in the statement~\ref{comment4} hold for various other geometries, 
such as $\brc{n(y)=1+cy,\,b(y)=b_{\rm c}}$, $\brc{n(y)=1,\,b(y)=b_{\rm c}\brkt{2+\sin(cy)}}$, and so on. 

The ambiguity in introducing $\Lct$ mentioned at the beginning of Sec.~\ref{cutoff_dep} does not affect most of the above conclusions. 
In fact, this ambiguity does not change the coefficients of $\cO(\Lct^5)$ and $\cO(\Lct^4)$. 
As for the statement~\ref{comment3}, it changes some numerical numbers in \eqref{warped_case:tt} and \eqref{warped_case:yy}, 
but does not change the $y$-dependence. 
Thus the required mass relations are affected, or the cancellation of the coefficients only happens in a specific way of introducing $\Lct$. 
Further studies are necessary to clarify whether such specific introduction of $\Lct$ that realizes the cancellation exists or not. 
We will leave this for future works. 

As noted in the footnote~\ref{fn:renormalizability}, the theory considered here is renormalizable because interaction terms are neglected in our setup.  
Hence all the contributions with positive powers in $\Lct$ should be absorbed into some parameters. 
The contributions to $\Lvev{T_{MN}}$ that have the same $y$-dependence as $g_{MN}$ can be absorbed to the 5D cosmological constant~$\Lmd_{\rm cc}^{\rm (5D)}$. 
However ,the results in Sec.~\ref{total_EMT} show that the cosmological constant term is not enough to renormalize all the divergent contributions. 
Therefore, we expect that such contributions are absorbed to some higher order terms in the background curvature terms, 
such as $\cR^{MNLP}\cR_{MNLP}$ or $\cR^{MN}\cR_{MN}$, which will be induced at loop level. 
In fact, in the case of $\brc{n(y)=1+cy^2,\;b(y)=b_{\rm c}}$, $C_{1}^{(tt)}(y)$ and $C_{1}^{(yy)}(y)$ contain terms proportional to 
$(1+cy^2)^{-1}$ and $(1+cy^2)^{-2}$. 
Such terms can arise from $\cR^{MNLP}\cR_{MNLP}$, $\cR^{MN}\cR_{MN}$ or $\cR^2$ as shown in \eqref{R^2-terms}. 

We should also note that no $\Lct^4$- and $\Lct^2$-contributions would appear in the momentum cutoff regularization, 
instead of the point-splitting regularization we have used. 
Therefore, such terms may be understood as artificial ones due to an inappropriate choice of the regularization. 
There may be  a specific choice of introducing $\Lct$ in the expressions of $\Lvev{T_{MN}}$
for the ambiguity mentioned at the beginning of Sec.~\ref{cutoff_dep}, 
with which the $\Lct^4$- and $\Lct^2$-contributions will disappear. 

In contrast to the previous works, we have not adopt the requirement~\eqref{requirement}, whose physical meaning is unclear. 
As a result, we obtain \eqref{finite_result} in the case of the flat background. 
This has a different radion ($b_{\rm c}$) dependence from the conventional result~\eqref{cv_result:flat}, 
and thus, it affects the modulus stabilization. 

The statement~\ref{comment4} indicates that 
$\Lvev{T_{MN}}$ have large $\cO(\Lct^3)$ values in a general background other than AdS or flat spacetime. 
From the viewpoint of the energy cost, it seems natural that the geometry of the extra dimension is forced 
to be the flat (or AdS) space, depending on the mass relations, through the time evolution governed by \eqref{Einstein_eq}. 
In other words, the dynamics hides the explicit $\Lct$-dependence of the vacuum energy.  
If this is the case, the geometry is determined by the mass spectrum, and the quantum contribution to the vacuum energy 
affects the modulus (radion) stabilization. 
It is interesting to check whether such process actually happens. 
We will discuss this issue by numerically solving the Einstein field equation~\eqref{Einstein_eq}  
in the subsequent papers.

\appendix

\section{Gamma matrices and covariant derivatives} \label{notations}
The 5D gamma matrices are
\begin{align}
 \Gm^0 &= \begin{pmatrix} & -\id_2 \\ -\id_2 & \end{pmatrix}, \;\;\;\;\;
 \Gm^{\udl{i}} = \begin{pmatrix} & \tau^i \\ -\tau^i & \end{pmatrix}, \;\;\;\;\;
 \Gm^4 = \begin{pmatrix} i\id_2 & \\ & -i\id_2 \end{pmatrix}, 
\end{align}
where $\udl{i}=1,2,3$ and $\tau^i$ ($i=1,2,3$) are the Pauli matrices. 
These satisfy
\begin{align}
 \brc{\Gm^A,\Gm^B} &= -2\eta^{AB}, 
\end{align}
where $\eta^{AB}=\diag(-1,1,1,1,1)$. 

Thus, the covariant derivatives for fermions under the metric ansatz~\eqref{metric_ansatz} are 
\begin{align}
 \cD_t &= \der_t-\frac{in'}{2b}\begin{pmatrix} & -\id_2 \\ \id_2 & \end{pmatrix}, \nonumber\\
 \cD_i &= \der_i+\frac{ia'}{2b}\begin{pmatrix} & \tau^i \\ \tau^i & \end{pmatrix}  
 = \der_i+\frac{i\asc n'}{2b}\begin{pmatrix} & \tau^i \\ \tau^i & \end{pmatrix}, \nonumber\\
 \cD_y &= \der_y. 
\end{align}

\section{Riemann tensor and Einstein equations} \label{Einstein_eqs}
\subsection{Riemann and Ricci tensors}
Under the metric ansatz~\eqref{metric_ansatz}, the nonvanishing components of the Riemann tensor are 
\begin{align}
 \cR_{titj} &=\brkt{-a\ddot{a}+\frac{\dot{n}a\dot{a}}{n}+\frac{nn'aa'}{b^2}}\dlt_{ij}, \nonumber\\
 \cR_{tijy} &= \brkt{a\dot{a}'-\frac{n'a\dot{a}}{n}-\frac{aa'\dot{b}}{b}}\dlt_{ij}, \nonumber\\
 \cR_{tyty} &= -b\ddot{b}+\frac{\dot{n}b\dot{b}}{n}+nn''-\frac{nn'b'}{b}, \nonumber\\
 \cR_{ijkl} &= a^2\brkt{\frac{\dot{a}^2}{n^2}-\frac{a^{\prime 2}}{b^2}}\brkt{\dlt_{ik}\dlt_{jl}-\dlt_{il}\dlt_{jk}}, \nonumber\\
 \cR_{yiyj} &= \brkt{-aa''+\frac{aa'b'}{b}+\frac{a\dot{a}b\dot{b}}{n^2}}\dlt_{ij}, 
\end{align}
and components related to these by the index symmetries, and the nonvanishing components of the Ricci tensor are 
\begin{align}
 \cR_{tt} &= -\frac{3\ddot{a}}{a}-\frac{\ddot{b}}{b}+\brkt{\frac{3\dot{a}}{a}+\frac{\dot{b}}{b}}\frac{\dot{n}}{n}
 +\frac{nn''}{b^2}+\brkt{\frac{3a'}{a}-\frac{b'}{b}}\frac{nn'}{b^2}, \nonumber\\
 \cR_{ty} &= -\frac{3\dot{a}'}{a}+\frac{3n'\dot{a}}{na}+\frac{3a'\dot{b}}{ab}, \nonumber\\
 \cR_{ij} &= \sbk{\frac{a\ddot{a}}{n^2}-\frac{aa''}{b^2}-\brkt{\frac{\dot{n}}{n}-\frac{2\dot{a}}{a}-\frac{\dot{b}}{b}}\frac{a\dot{a}}{n^2}
 -\brkt{\frac{n'}{n}+\frac{2a'}{a}-\frac{b'}{b}}\frac{aa'}{b^2}}\dlt_{ij}, \nonumber\\
 \cR_{yy} &= \frac{b\ddot{b}}{n^2}-\frac{n''}{n}-\frac{3a''}{a}-\brkt{\frac{\dot{n}}{n}-\frac{3\dot{a}}{a}}\frac{b\dot{b}}{n^2}
 +\brkt{\frac{n'}{n}+\frac{3a'}{a}}\frac{b'}{b}. 
 \label{comp:Ricci_tensor}
\end{align}

\subsection{Einstein equations} \label{Einstein_eqs:2}
\ignore{
Under the metric ansatz~\eqref{metric_ansatz}, the nonvanishing components of the Einstein tensor~$G_{MN}\equiv \cR_{MN}-\frac{1}{2}g_{MN}\cR$ are
\begin{align}
 G_{tt} &= \frac{3\dot{a}}{a}\brkt{\frac{\dot{a}}{a}+\frac{\dot{b}}{b}}
 -\frac{3n^2}{b^2}\sbk{\frac{a''}{a}+\frac{a'}{a}\brkt{\frac{a'}{a}-\frac{b'}{b}}}, \nonumber\\
 G_{ty} &= -\frac{3\dot{a}'}{a}+\frac{3n'\dot{a}}{na}+\frac{3a'\dot{b}}{ab}, \nonumber\\
 G_{ij} &= \frac{\dlt_{ij}a^2}{n^2}\brkt{-\frac{2\ddot{a}}{a}-\frac{\ddot{b}}{b}-\frac{\dot{a}^2}{a^2}
 +\frac{2\dot{n}\dot{a}}{na}+\frac{\dot{n}\dot{b}}{nb}-\frac{2\dot{a}\dot{b}}{ab}} \nonumber\\
 &\quad
 +\frac{\dlt_{ij}a^2}{b^2}\brkt{\frac{n''}{n}+\frac{2a''}{a}+\frac{a^{\prime 2}}{a^2}+\frac{2n'a'}{na}-\frac{n'b'}{nb}-\frac{2a'b'}{ab}}, \nonumber\\
 G_{yy} &= -\frac{3b^2}{n^2}\brkt{\frac{\ddot{a}}{a}+\frac{\dot{a}^2}{a^2}-\frac{\dot{n}\dot{a}}{na}}
 +\frac{3a^{\prime 2}}{a^2}+\frac{3n'a'}{na}. 
\end{align}
}
Utilizing the expressions in \eqref{comp:Ricci_tensor}, the Einstein equation~\eqref{Einstein_eq} is summarized as
\begin{align}
 \frac{\ddot{a}}{a} &= \frac{\dot{a}}{a}\brkt{\frac{\dot{n}}{n}-\frac{\dot{a}}{a}}
 +\frac{n^2}{b^2}\frac{a'}{a}\brkt{\frac{n'}{n}+\frac{a'}{a}}-\frac{\kp_5n^2}{3b^2}\Lvev{T_{yy}}+\frac{n^2}{3}\Lmd_5, \nonumber\\
 \frac{\ddot{b}}{b} &= \frac{4\dot{a}^2}{a^2}+\frac{\dot{b}}{b}\brkt{\frac{\dot{n}}{n}+\frac{\dot{a}}{a}}
 +\frac{n^2}{b^2}\sbk{\frac{n''}{n}-\frac{a''}{a}-\frac{4a^{\prime 2}}{a^2}-\frac{b'}{b}\brkt{\frac{n'}{n}-\frac{a'}{a}}} \nonumber\\
 &\quad
 -\kp_5\brkt{\Lvev{T_{tt}}+\frac{n^2}{a^2}\Lvev{T_{\rm 3d}}-\frac{2n^2}{3b^2}\Lvev{T_{yy}}}-\frac{2n^2}{3}\Lmd_5, 
 \label{evolution_eqs}
\end{align}
where $\Lvev{T_{ij}}\equiv \dlt_{ij}\Lvev{T_{\rm 3d}}$, 
with 
\begin{align}
 \frac{\dot{a}}{a}\brkt{\frac{\dot{a}}{a}+\frac{\dot{b}}{b}}
 -\frac{n^2}{b^2}\sbk{\frac{a''}{a}+\frac{a'}{a}\brkt{\frac{a'}{a}-\frac{b'}{b}}} 
 &= \frac{\kp_5}{3}\Lvev{T_{tt}}+\frac{n^2}{3}\Lmd_5, \nonumber\\
 \frac{\dot{a}'}{a} &= \frac{n'\dot{a}}{na}+\frac{a'\dot{b}}{ab}. 
 \label{constraints}
\end{align}
The equations in \eqref{constraints} do not contain the second time derivatives, and thus 
should be understood as constraints on the initial conditions.\footnote{
Once the equations in \eqref{constraints} are satisfied at the initial time, they remain to be hold at later times. 
}

If we restrict the metric components as \eqref{metric_assumption}, the above equations become
\begin{align}
 \frac{\ddot{n}}{n} &= -\frac{\dot{a}_{\rm sc}}{\asc}\brkt{\frac{3\dot{n}}{n}+\frac{\dot{a}_{\rm sc}}{\asc}}-\frac{\ddot{a}_{\rm sc}}{\asc}
 +\frac{2n^{\prime 2}}{b^2}-\frac{\kp_5n^2}{3b^2}\Lvev{T_{yy}}+\frac{n^2}{3}\Lmd_5, \nonumber\\
 \frac{\ddot{b}}{b} &= 4\brkt{\frac{\dot{n}}{n}+\frac{\dot{a}_{\rm sc}}{\asc}}^2
 +\frac{\dot{b}}{b}\brkt{\frac{2\dot{n}}{n}+\frac{\dot{a}_{\rm sc}}{\asc}}
 -\frac{4n^{\prime 2}}{b^2}+\frac{2\kp_5n^2}{3b^2}\Lvev{T_{yy}}-\frac{2n^2}{3}\Lmd_5, 
 \label{evolve_eq:restrict}
\end{align}
with 
\begin{align}
 \brkt{\frac{\dot{n}}{n}+\frac{\dot{a}_{\rm sc}}{\asc}}\brkt{\frac{\dot{n}}{n}+\frac{\dot{a}_{\rm sc}}{\asc}+\frac{\dot{b}}{b}}
 -\frac{n^2}{b^2}\sbk{\frac{n''}{n}+\frac{n'}{n}\brkt{\frac{n'}{n}-\frac{b'}{b}}} 
 &= \frac{\kp_5}{3}\Lvev{T_{tt}}+\frac{n^2}{3}\Lmd_5, 
 \label{constraint:restrict1}
\end{align}
and
\begin{align}
 \frac{\dot{n}'}{n} &= \frac{n'}{n}-\frac{n'\dot{a}_{\rm sc}}{n\asc}+\brkt{\frac{\dot{n}}{n}+\frac{\dot{a}_{\rm sc}}{\asc}}^{-1}\frac{n'\dot{b}}{nb}. 
 \label{constraint:restrict2}
\end{align}

\subsection{Higher curvature terms}
Under the assumption~\eqref{metric_assumption} and neglecting the $t$-dependence of the background, we have
\begin{align}
 \cR^{MNLP}\cR_{MNLP} &= \frac{8}{b^4}\sbk{\frac{3n^{\prime 4}}{n^4}+2\brkt{\frac{n''}{n}-\frac{n'b'}{nb}}^2}, \nonumber\\
 \cR^{MN}\cR_{MN} &= \frac{4}{b^4}\sbk{5\brkt{\frac{n''}{n}-\frac{n'b'}{nb}}^2
 +\frac{6n^{\prime 2}}{n^2}\brkt{\frac{n''}{n}-\frac{n'b'}{nb}}+\frac{9n^{\prime 4}}{n^4}}, \nonumber\\
 \cR^2 &= \frac{64}{b^4}\brkt{\frac{n''}{n}-\frac{n'b'}{nb}+\frac{3n^{\prime 2}}{n^2}}^2. 
 \label{R^2-terms}
\end{align}

\section{Basis functions} \label{basis_fcns}
\subsection{Scalar sector} \label{scalar:basis_fct}
Two independent solutions of \eqref{EOM:tlPhi} for a fixed value of $\pf^2$ can be chosen as $C_0(y;\pf^2)$ and $S_0(y;\pf^2)$ that satisfy 
\begin{align}
 C_0(0;\pf^2) &= 1, \;\;\;\;\;
 C'_0(0;\pf^2) = 0, \nonumber\\
 S_0(0;\pf^2) &= 0, \;\;\;\;\;
 S'_0(0;\pf^2) = 1. 
 \label{BC0:scalar}
\end{align}
A general solution of \eqref{EOM:tlPhi} can be expressed as a linear combination of these. 
So we refer to them as basis functions. 
They also satisfy the Wronskian relation, 
\begin{align}
 \cW_0(y) &\equiv C_0(y;\pf^2)S_0'(y;\pf^2)-S_0(y;\pf^2)C_0'(y;\pf^2) = \frac{n^4(0)b(y)}{b(0)n^4(y)}. 
 \label{def:cW_0}
\end{align}

We can also choose another set of basis functions, $C_L(y;\pf^2)$ and $S_L(y;\pf^2)$ that satisfy 
\begin{align}
 C_L(L;\pf^2) &= 1, \;\;\;\;\;
 C_L'(L;\pf^2) = 0, \nonumber\\
 S_L(L;\pf^2) &= 0, \;\;\;\;\;
 S_L'(L;\pf^2) = 1. 
 \label{BCL:scalar}
\end{align}
The Wronskian relation for these is 
\begin{align}
 \cW_L(y) &\equiv C_L(y;\pf^2)S_L'(y;\pf^2)-S_L(y;\pf^2)C_L'(y;\pf^2) = \frac{n^4(L)b(y)}{b(L)n^4(y)}. 
\end{align}

The relations between the two sets of the basis functions are given by
\begin{align}
 C_L(y;\pf^2) &= \frac{b(0)n^4(L)}{n^4(0)b(L)}\brc{S_0'(L;\pf^2)C_0(y;\pf^2)-C_0'(L;\pf^2)S_0(y;\pf^2)}, \nonumber\\
 S_L(y;\pf^2) &= \frac{b(0)n^4(L)}{n^4(0)b(L)}\brc{C_0(L;\pf^2)S_0(y;\pf^2)-S_0(L;\pf^2)C_0(y;\pf^2)}. 
 \label{rel:CS_0L}
\end{align}

\subsection{Spinor sector} \label{spinor:basis_fct}
Two independent solutions of 
\begin{align}
 \cO_\pm f^{(\pm)}(y;\pf^2) &= 0, 
\end{align}
where $\cO_\pm$ is defined in \eqref{def:cO_pm}, 
can be chosen as $C_0^{(\pm)}(y;\pf^2)$ and $S_0^{(\pm)}(y;\pf^2)$ that satisfy
\begin{align}
 C_0^{(\pm)}(0;\pf^2) &= 1, \;\;\;\;\;
 C_0^{(\pm)\prime}(0;\pf^2) = 0, \nonumber\\
 S_0^{(\pm)}(0;\pf^2) &= 0, \;\;\;\;\;
 S_0^{(\pm)\prime}(0;\pf^2) = 1. 
 \label{BC0:spinor}
\end{align}
These are the basis functions in the fermionic sector. 
They satisfy the Wronskian relation, 
\begin{align}
 \cW_0^{(\pm)}(y) &\equiv C_0^{(\pm)}(y;\pf^2)S_0^{(\pm)\prime}(y;\pf^2)-S_0^{(\pm)}(y;\pf^2)C_0^{(\pm)\prime}(y;\pf^2) 
 = \frac{n(0)b(y)}{b(0)n(y)}. 
\end{align}

We can also choose another set of basis functions, $C_L^{(\pm)}(y;\pf^2)$ and $S_L^{(\pm)}(y;\pf^2)$ that satisfy 
\begin{align}
 C_L^{(\pm)}(L;\pf^2) &= 1, \;\;\;\;\;
 C_L^{(\pm)\prime}(L;\pf^2) = 0, \nonumber\\
 S_L^{(\pm)}(L;\pf^2) &= 0, \;\;\;\;\;
 S_L^{(\pm)\prime}(L;\pf^2) = 1. 
 \label{BCL:spinor}
\end{align}
The Wronskian relation for these is 
\begin{align}
 \cW_L^{(\pm)}(y) &\equiv C_L^{(\pm)}(y;\pf^2)S_L^{(\pm)\prime}(y;\pf^2)-S_L^{(\pm)}(y;\pf^2)C_L^{(\pm)\prime}(y;\pf^2) 
 = \frac{n(L)b(y)}{b(L)n(y)}. 
\end{align}

\subsection{Simple examples}
In the flat geometry, i.e., $\brc{n(y)=1,\,b(y)=b_{\rm c}}$ ($b_{\rm c}$ : constant), 
the basis functions for the scalar and the spinor sectors become the same, and are given by~\footnote{
For $\pf^2+M^2<0$ (i.e., $p_t^2>\pth^2+M^2$), they are expressed as
\begin{align}
 C_u(y;\pf^2) &= \cos\brkt{b_{\rm c}\sqrt{\abs{\pf^2+M^2}}(y-u)}, \;\;\;\;\;
 S_u(y;\pf^2) = \frac{\sin\brkt{b_{\rm c}\sqrt{\abs{\pf^2+M^2}}(y-u)}}{b_{\rm c}\sqrt{\abs{\pf^2+M^2}}}. 
\end{align}
}
\begin{align}
 C_u(y;\pf^2) &= C_u^{(\pm)}(y;\pf^2) = \cosh\brc{b_{\rm c}\sqrt{\pf^2+M^2}(y-u)}, \nonumber\\
 S_u(y;\pf^2) &= S_u^{(\pm)}(y;\pf^2) = \frac{\sinh\brc{b_{\rm c}\sqrt{\pf^2+M^2}(y-u)}}{b_{\rm c}\sqrt{\pf^2+M^2}}, 
\end{align}
where $u=0,L$. 

In the warped geometry, i.e., $\brc{n(y)=e^{-ky},\,b(y)=b_{\rm c}}$, the basis functions for the scalar sector are 
\begin{align}
 C_u(y;\pf^2) &= e^{2k(y-u)}\sbk{A_u(\pf^2) I_\nu\brkt{e^{ky}\frac{b_{\rm c}\sqrt{\pf^2}}{k}}+B_u(\pf^2) K_\nu\brkt{e^{ky}\frac{b_{\rm c}\sqrt{\pf^2}}{k}}}, \nonumber\\
 S_u(y;\pf^2) &= \frac{e^{2k(y-u)}}{k}\left[K_\nu\brkt{e^{ku}\frac{b_{\rm c}\sqrt{\pf^2}}{k}}I_\nu\brkt{e^{ky}\frac{b_{\rm c}\sqrt{\pf^2}}{k}} \right.\nonumber\\
 &\hspace{22mm}\left.
 -I_\nu\brkt{e^{ku}\frac{b_{\rm c}\sqrt{\pf^2}}{k}}K_\nu\brkt{e^{ky}\frac{b_{\rm c}\sqrt{\pf^2}}{k}}\right], 
\end{align}
where $\nu=\sqrt{4+b_{\rm c}^2M^2/k^2}$, the functions~$I_\nu(z)$ and $K_\nu(z)$ are the modified Bessel functions, and 
\begin{align}
 A_u(\pf^2) &\equiv -(2+\nu)K_\nu\brkt{e^{ku}\frac{b_{\rm c}\sqrt{\pf^2}}{k}}
 +e^{ku}\frac{b_{\rm c}\sqrt{\pf^2}}{k}K_{\nu+1}\brkt{e^{ku}\frac{b_{\rm c}\sqrt{\pf^2}}{k}}, \nonumber\\
 B_u(\pf^2) &\equiv (2+\nu)I_\nu\brkt{e^{ku}\frac{b_{\rm c}\sqrt{\pf^2}}{k}}
 +e^{ku}\frac{b_{\rm c}\sqrt{\pf^2}}{k}I_{\nu+1}\brkt{e^{ku}\frac{b_{\rm c}\sqrt{\pf^2}}{k}}. 
\end{align}
Those for the spinor sector are 
\begin{align}
 C_u^{(\pm)}(y;\rho^2) &= e^{\frac{k}{2}(y-u)}\sbk{A_u(\pf^2)I_\nu\brkt{e^{ky}\frac{b_{\rm c}\sqrt{\pf^2}}{k}}
 +B_u(\pf^2)K_\nu\brkt{e^{ky}\frac{b_{\rm c}\sqrt{\pf^2}}{k}}}, \nonumber\\
 S_u^{(\pm)}(y;\rho^2) &= \frac{e^{\frac{k}{2}(y-u)}}{k}
 \left\{K_\nu\brkt{e^{ku}\frac{b_{\rm c}\sqrt{\pf^2}}{k}}I_\nu\brkt{e^{ky}\frac{b_{\rm c}\sqrt{\pf^2}}{k}} \right.\nonumber\\
 &\hspace{22mm}\left.
 -I_\nu\brkt{e^{ku}\frac{b_{\rm c}\sqrt{\pf^2}}{k}}K_\nu\brkt{e^{ky}\frac{b_{\rm c}\sqrt{\pf^2}}{k}}\right\}, 
\end{align}
where $\nu=\frac{1}{2}\pm\frac{b_{\rm c}M}{k}$, and 
\begin{align}
 A_u(\pf^2) &\equiv -\brkt{\frac{1}{2}+\nu}K_\nu\brkt{e^{ku}\frac{b_{\rm c}\sqrt{\pf^2}}{k}}
 +e^{ku}\frac{b_{\rm c}\sqrt{\pf^2}}{k}K_{\nu+1}\brkt{e^{ku}\frac{b_{\rm c}\sqrt{\pf^2}}{k}}, \nonumber\\
 B_u(\pf^2) &\equiv \brkt{\frac{1}{2}+\nu}I_\nu\brkt{e^{ku}\frac{b_{\rm c}\sqrt{\pf^2}}{k}}
 +e^{ku}\frac{b_{\rm c}\sqrt{\pf^2}}{k}I_{\nu+1}\brkt{e^{ku}\frac{b_{\rm c}\sqrt{\pf^2}}{k}}. 
\end{align}

\section{5D propagator} \label{derivation:5D_propagator}
\subsection{Scalar sector} \label{5D_propagator:scalar_sector}
Here we derive the expression of the 5D propagator~$\tl{G}_{\rm B}(p_\mu,y,y')$, which is defined by 
\begin{align}
 \langle 0|T\Phi(x^\mu,y)\Phi(x^{\prime\nu},y')|0\rangle 
 &= \int\frac{d^4p}{i(2\pi)^4}\;e^{ip\cdot(x-x')}\tl{G}_{\rm B}(p_\mu,y,y'). 
 \label{def:tlG_B}
\end{align}
It is obtained as a solution of 
\begin{align}
 \brc{\frac{\pf^2}{n^2(y)}
 -\frac{1}{b^2(y)}\sbk{\der_y^2+\brkt{\frac{4n'(y)}{n(y)}-\frac{b'(y)}{b(y)}}\der_y}+M^2}\tl{G}_{\rm B}(p_\mu,y,y') 
 &= \frac{\dlt(y-y')}{\sqrt{-g(y)}} \nonumber\\
 &= \frac{\dlt(y-y')}{\asc^3n^4(y)b(y)}. 
 \label{eq:tlG_B}
\end{align}
By definition, this satisfies
\begin{align}
 \tl{G}_{\rm B}(p_\mu,y,y') &= \tl{G}_{\rm B}(p_\mu,y',y). 
 \label{rel:tlG_B:1}
\end{align}
From the boundary conditions in \eqref{BC:scalar}, $\tl{G}_B(p_\mu,y,y')$ also satisfies
\begin{align}
 \der_y \tl{G}_{\rm B}(p_\mu,0,y') &= \der_y\tl{G}_{\rm B}(p_\mu,L,y') = 0. 
 \label{BC:tlG_B}
\end{align}
By integrating \eqref{eq:tlG_B} over infinitesimal interval that contain $y=y'$, we obtain
\begin{align}
 \lim_{y\to y'}\sbk{\der_y\tl{G}_{\rm B}(p_\mu,y,y')-\der_y\tl{G}_{\rm B}(p_\mu,y,y')} &= -\frac{b(y')}{\asc^3n^4(y')}. 
 \label{rel:tlG_B:2}
\end{align}

From \eqref{BC:tlG_B}, we can express $\tl{G}_B(p_\mu,y,y')$ in terms of the basis functions in Appendix~\ref{scalar:basis_fct} as
\begin{align}
 \tl{G}_{\rm B}(p_\mu,y,y') &= \vth(y-y')C_L(y;\pf^2)\alp_{>}(y';\pf^2)+\vth(y'-y)C_0(y;\pf^2)\alp_<(y';\pf^2), 
\end{align}
where $\vth(y)$ is the Heaviside step function. 
By using \eqref{rel:tlG_B:1} and \eqref{rel:tlG_B:2}, we find that
\begin{align}
 \alp_{>}(y';\pf^2) &= -\frac{b(y')C_0(y';\pf^2)}{\asc^3n^4(y')\brkt{C_0C_L'-C_0'C_L}(y';\pf^2)} 
 = \frac{b(L)C_0(y';\pf^2)}{\asc^3n^4(L)C_0'(L;\pf^2)}, \nonumber\\
 \alp_{<}(y';\pf^2) &= -\frac{b(y')C_L(y';\pf^2)}{\asc^3n^4(y')\brkt{C_0C_L'-C_0'C_L}(y';\pf^2)} 
 = \frac{b(L)C_L(y';\pf^2)}{\asc^3n^4(L)C_0'(L;\pf^2)}. 
\end{align}
At the second equalities, we have used that
\begin{align}
 \brkt{C_0C_L'-C_0'C_L}(y';\pf^2) &= -C_0'(L;\pf^2)\frac{n^4(L)b(y')}{b(L)n^4(y')}, 
\end{align}
which follows from \eqref{def:cW_0} and \eqref{rel:CS_0L}. 
Note that $\tl{G}_{\rm B}(p_\mu,y,y')$ is a function of $\pf^2$. 
Thus it is denoted as $\tl{G}_{\rm B}(\pf^2,y,y')$ in the main text.

\subsection{Spinor sector}\label{5Dprop:spinor}
In the 2-component spinor notation, the 5D propagator is written as
\begin{align}
 \vev{T\Psi(x^\mu,y)\bar{\Psi}(x^{\prime \nu},y')} &= 
 -\begin{pmatrix} \vev{T\chi(x^\mu,y)\bar{\lmd}^\dagger(x^{\prime\nu},y')} & \vev{T\chi(x^\mu,y)\chi^\dagger(x^{\prime\nu},y')} \\
 \vev{T\bar{\lmd}(x^\mu,y)\bar{\lmd}^\dagger(x^{\prime \nu},y')} & \vev{T\bar{\lmd}(x^\mu,y)\chi^\dagger(x^{\prime\nu},y')} \end{pmatrix}. 
\end{align}
We move to the 4D momentum basis. 
\begin{align}
 \vev{T\chi(x^\mu,y)\bar{\lmd}^\dagger(x^{\prime\nu},y')} &\equiv \int\frac{d^4p}{i(2\pi)^4}\;e^{ip\cdot(x-x')}\frac{\hat{G}_{\chi\lmd}(p_\mu,y,y')}{n^2(y)n^2(y')}, 
 \nonumber\\
 \vev{T\chi(x^\mu,y)\chi^\dagger(x^{\prime\nu},y')} &\equiv \int\frac{d^4p}{i(2\pi)^4}\;e^{ip\cdot(x-x')}\frac{\hat{G}_{\chi\chi}(p_\mu,y,y')}{n^2(y)n^2(y')}, 
 \nonumber\\
 \vev{T\bar{\lmd}(x^\mu,y)\bar{\lmd}^\dagger(x^{\prime \nu},y')} &\equiv \int\frac{d^4p}{i(2\pi)^4}\;e^{ip\cdot(x-x')}
 \frac{\hat{G}_{\lmd\lmd}(p_\mu,y,y')}{n^2(y)n^2(y')}, \nonumber\\
 \vev{T\bar{\lmd}(x^\mu,y)\chi^\dagger(x^{\prime\nu},y')} &\equiv \int\frac{d^4p}{i(2\pi)^4}\;e^{ip\cdot(x-x')}
 \frac{\hat{G}_{\lmd\chi}(p_\mu,y,y')}{n^2(y)n^2(y')}. 
 \label{def:hatGs}
\end{align}
Then, $\hat{G}_{ab}(p_\mu,y,y')$ ($a,b=\chi,\lmd$) are obtained as solutions of
\begin{align}
 \begin{pmatrix} \frac{1}{b}\brkt{\der_y-bM} & -\frac{1}{n}\brkt{p_t-\frac{1}{\asc}\tau^ip_i} \\
 -\frac{1}{n}\brkt{p_t+\frac{1}{\asc}\tau^ip_i} & -\frac{1}{b}\brkt{\der_y+bM} \end{pmatrix}\hat{G}_{\rm F}(p_\mu,y,y')
 &= \frac{n^2(y)n^2(y')\dlt(y-y')}{\sqrt{-g}}\id_4 \nonumber\\
 &= \frac{\dlt(y-y')}{\asc^3b(y)}\id_4, 
\end{align}
where
\begin{align}
 \hat{G}_{\rm F}(p_\mu,y,y') &\equiv -\begin{pmatrix} \hat{G}_{\chi\lmd}(p_\mu,y,y') & \hat{G}_{\chi\chi}(p_\mu,y,y') \\
 \hat{G}_{\lmd\lmd}(p_\mu,y,y') & \hat{G}_{\lmd\chi}(p_\mu,y,y') \end{pmatrix}. 
 \label{def:hatG_F}
\end{align}
By definition, $\hat{G}_{ab}(p_\mu,y,y')$ ($a,b=\chi,\lmd$) satisfy
\begin{align}
 \hat{G}_{\chi\lmd}^\dagger(p_\mu,y,y') &= \hat{G}_{\lmd\chi}(p_\mu,y',y), \nonumber\\
 \hat{G}_{\chi\chi}^\dagger(p_\mu,y,y') &= \hat{G}_{\chi\chi}(p_\mu,y',y), \nonumber\\
 \hat{G}_{\lmd\lmd}^\dagger(p_\mu,y,y') &= \hat{G}_{\lmd\lmd}(p_\mu,y',y), \nonumber\\
 \hat{G}_{\lmd\chi}^\dagger(p_\mu,y,y') &= \hat{G}_{\chi\lmd}(p_\mu,y',y). 
\end{align}
From the boundary conditions~\eqref{BC:chi} and \eqref{BC:lmd}, $\hat{G}_{ab}(p_\mu,y,y')$ also satisfy
\begin{align}
 \hat{G}_{\chi\lmd}(p_\mu,0,y') &= \hat{G}_{\chi\lmd}(p_\mu,L,y') = 0, \nonumber\\
 \hat{G}_{\chi\chi}(p_\mu,0,y') &= \hat{G}_{\chi\chi}(p_\mu,L,y') = 0, 
\end{align}
and
\begin{align}
 \left.\brkt{\der_y+bM}\hat{G}_{\lmd\lmd}\right|_{y=0} &= \left.\brkt{\der_y+bM}\hat{G}_{\lmd\lmd}\right|_{y=L} = 0, \nonumber\\
 \left.\brkt{\der_y+bM}\hat{G}_{\lmd\chi}\right|_{y=0} &= \left.\brkt{\der_y+bM}\hat{G}_{\lmd\chi}\right|_{y=L} = 0. 
\end{align}

Following a similar procedure to the scalar case, we obtain
\begin{align}
 \hat{G}_{\chi\chi}(p_\mu,y,y') &= -\frac{n(L)}{\asc^3b(L)S_L^{(-)}(0;\pf^2)}\brkt{p_t\id_2-\frac{p_i}{\asc}\tau^i} \nonumber\\
 &\hspace{5mm}
 \times\brc{\vth(y-y')H_{L\chi}(y;\pf^2)H_{0\chi}(y';\pf^2)+\vth(y'-y)H_{0\chi}(y;\pf^2)H_{L\chi}(y';\pf^2)}, \nonumber\\
 \hat{G}_{\chi\lmd}(p_\mu,y,y') &= -\frac{n(L)}{\asc^3b(L)S_L^{(-)}(0;\pf^2)}\id_2 \nonumber\\
 &\hspace{5mm}
 \times\brc{\vth(y-y')H_{L\chi}(y;\pf^2)H_{0\lmd}(y';\pf^2)+\vth(y'-y)H_{0\chi}(y;\pf^2)H_{L\lmd}(y';\pf^2)}, \nonumber\\
 \hat{G}_{\lmd\chi}(p_\mu,y,y') &= -\frac{n(L)}{\asc^3b(L)S_L^{(-)}(0;\pf^2)}\id_2 \nonumber\\
 &\hspace{5mm}
 \times\brc{\vth(y-y')H_{L\lmd}(y;\pf^2)H_{0\chi}(y';\pf^2)+\vth(y'-y)H_{0\lmd}(y;\pf^2)H_{L\chi}(y';\pf^2)}, \nonumber\\
 \hat{G}_{\lmd\lmd}(p_\mu,y,y') &= \frac{n(L)}{\asc^3b(L)S_L^{(-)}(0;\pf^2)}\frac{1}{\pf^2}\brkt{p_t\id_2+\frac{p_i}{\asc}\tau^i} \nonumber\\
 &\hspace{5mm}
 \times\brc{\vth(y-y')H_{L\lmd}(y;\pf^2)H_{0\lmd}(y';\pf^2)+\vth(y'-y)H_{0\lmd}(y;\pf^2)H_{L\lmd}(y';\pf^2)}, 
 \label{expr:hatGs}
\end{align}
where
\begin{align}
 H_{0\chi}(y;\pf^2) &\equiv \frac{b(0)}{n(0)}S_0^{(-)}(y;\pf^2), \nonumber\\
 H_{0\lmd}(y;\pf^2) &\equiv C_0^{(+)}(y;\pf^2)-b(0)MS_0^{(+)}(y;\pf^2), \nonumber\\
 H_{L\chi}(y;\pf^2) &\equiv \frac{b(L)}{n(L)}S_L^{(-)}(y;\pf^2), \nonumber\\
 H_{L\lmd}(y;\pf^2) &\equiv C_L^{(+)}(y;\pf^2)-b(L)MS_L^{(+)}(y;\pf^2). 
 \label{def:H0L}
\end{align}
The basis functions~$C_{0,L}^{(\pm)}(y;\pf^2)$ and $S_{0,L}^{(\pm)}(y;\pf^2)$ are defined in Appendix~\ref{spinor:basis_fct}.

\section{Expansions for large momentum} \label{LM_expansion}
Here we derive approximate expressions of the basis functions and the 5D propagators for large 4-momentum~$\pf$ or $\rho$. 
We focus on the case that $y<y'$, and $\rho\geq\bar{\rho}>0$. 
\subsection{Scalar sector} \label{apx_expr:boson}
\subsubsection{Basis functions}
The basis functions~$C_u(y;\rho^2)$ and $S_u(y;\rho^2)$ ($u=0,L$) are solutions of 
\begin{align}
 \sbk{\der_y^2+\brkt{\frac{4n'}{n}-\frac{b'}{b}}\der_y-\frac{b^2}{n^2}\rho^2-b^2M^2}f_u(y;\rho) &= 0. 
 \label{md_eq:scalar}
\end{align}
For large values of $\rho$, a solution~$f_u(y;\rho)$ of this equation can be expanded as
\begin{align}
 f_u(y;\rho) &= \frac{\exp\sbk{\rho\,\cU_u(y)}}{n^{3/2}(y)}
 \sbk{1+\frac{g_{u,1}(y)}{\rho}+\frac{g_{u,2}(y)}{\rho^2}+\frac{g_{u,3}(y)}{\rho^3}+\cdots}, 
 \label{def:f_u}
\end{align}
where 
\begin{align}
 \cU_u(y) &\equiv \int_u^y d\tl{y}\;\frac{b(\tl{y})}{n(\tl{y})}, 
\end{align}
and the functions~$g_{u,m}(y)$ ($m=1,2,3,\cdots$) are solutions of 
\begin{align}
 g_{u,1}' &= \frac{3n''}{4b}+\frac{9n^{\prime 2}}{8nb}-\frac{3n'b'}{4b^2}+\frac{nbM^2}{2}, \nonumber\\
 g_{u,m\geq 2}' &= -\frac{n}{2b}\sbk{g_{u,m-1}''+\brkt{\frac{n'}{n}-\frac{b'}{b}}g'_{u,m-1}
 -\brkt{\frac{3n''}{2n}+\frac{9n^{\prime 2}}{4n^2}-\frac{3n'b'}{2nb}+b^2M^2}g_{u,m-1}}. 
\end{align}

Note that when $f_u(y;\rho)$ is a solution of \eqref{md_eq:scalar}, 
$f_u(y;-\rho)$ is also a solution of it. 
Hence we can express the basis functions as linear combinations of them. 
Since the basis functions are functions of $\rho^2$, they should be proportional to $f_u(y;\rho)\pm f_u(y;-\rho)$. 
Taking into account the conditions in \eqref{BC0:scalar} and \eqref{BCL:scalar}, 
the basis function~$C_u(y;\rho^2)$ is found to be
\begin{align}
 C_u(y;\rho^2) &= \frac{n^{3/2}(u)}{2}\sbk{f_u(y;\rho)+f_u(y;-\rho)}, 
 \label{ap:C_u}
\end{align}
where the functions~$g_{u,m}(y)$ $(m=1,2,3,\cdots)$ are subject to the boundary conditions, 
\begin{align}
 g_1(u) &= \frac{3n'(u)}{2b(u)}, \nonumber\\
 g_{u,2l}(u) &= 0, \nonumber\\
 g_{2l+1}(u) &= -\frac{n(u)}{b(u)}g_{2l}'(u)+\frac{3n'(u)}{2b(u)}g_{2l}(u) = -\frac{n(u)}{b(u)}g_{2l}'(u). \;\;\;\;\; \brkt{l=1,2,3,\cdots}
\end{align}

As for the other basis function~$S_u(y;\rho^2)$, it is expressed as
\begin{align}
 S_u(y;\rho^2) &= \frac{n^{5/2}(u)}{2\rho b(u)}\sbk{f_u(y;\rho)-f_u(y;-\rho)}, 
\end{align}
where $g_{u,m}(y)$ $(m=1,2,3,\cdots)$ ar subject to 
\begin{align}
 g_{2l-1}(u) &= 0, \;\;\;\;\;
 g_{2l}(u) = -\frac{n(u)}{b(u)}g_{2l-1}'(u).  \;\;\;\;\; \brkt{l=1,2,3,\cdots}
\end{align}

\subsubsection{5D propagator}
Now we consider the behavior of the 5D propagator~$\tl{G}_{B<}(\rho^2,y,y')$ at large $\rho$. 
Note that
\begin{align}
 \cU_0(y) &= \int_0^y d\tl{y}\;\frac{b(\tl{y})}{n(\tl{y})} \geq 0, \;\;\;\;\;
 \cU_L(y') = \int_L^{y'} d\tl{y}\;\frac{b(\tl{y})}{n(\tl{y})} \leq 0. 
\end{align}
Thus, when $\bar{\rho}\abs{\cU_u(y)}\gg 1$, we have
\begin{align}
 C_0(y;\rho^2) &= \frac{n^{3/2}(0)}{2n^{3/2}(y)}e^{\rho\cU_0(y)}\sbk{1+\frac{g_{0,1}(y)}{\rho}+\frac{g_{0,2}(y)}{\rho^2}
 +\frac{g_{0,3}(y)}{\rho^3}+\cdots}, \nonumber\\
 C_L(y';\rho^2) &= \frac{n^{3/2}(L)}{2n^{3/2}(y')}e^{-\rho\cU_L(y')}
 \sbk{1-\frac{g_{L,1}(y')}{\rho}+\frac{g_{L,2}(y')}{\rho^2}-\frac{g_{L,3}(y')}{\rho^3}+\cdots}, \nonumber\\
 C_0'(L;\rho^2) &= \frac{\rho n^{3/2}(0)b(L)}{2n^{5/2}(L)}e^{\rho\cU_0(L)}
 \sbk{1+\frac{\cG_1}{\rho}+\frac{\cG_2}{\rho^2}+\frac{\cG_3}{\rho^3}+\cdots}, 
\end{align}
where 
\begin{align}
 \cG_1 &\equiv g_{0,1}(L)-\frac{3n'(L)}{2b(L)}, \nonumber\\
 \cG_2 &\equiv g_{0,2}(L)+\frac{n(L)}{b(L)}g_{0,1}'(L)-\frac{3n'(L)}{2b(L)}g_{0,1}(L), \nonumber\\
 \cG_3 &\equiv g_{0,3}(L)+\frac{n(L)}{b(L)}g_{0,2}'(L)-\frac{3n'(L)}{2b(L)}g_{0,2}(L), \nonumber\\
 &\vdots
\end{align}
Then, we have
\begin{align}
 \tl{G}_{\rm B<}(\rho^2,y,y') &= \frac{b(L)}{\asc^3 n^4(L)}\frac{C_0(y;\rho^2)C_L(y';\rho^2)}{C_0'(L;\rho^2)} \nonumber\\
 &\simeq \frac{e^{-\rho\Dlt(y,y')}}{2\rho\asc^3n^{3/2}(y)n^{3/2}(y')}
 \sbk{1+\frac{\cH_1(y,y')}{\rho}+\frac{\cH_2(y,y')}{\rho^2}+\frac{\cH_3(y,y')}{\rho^3}+\cdots}, 
 \label{tlG_B}
\end{align}
where
\begin{align}
 \Dlt(y,y') &\equiv -\cU_0(y)+\cU_L(y')+\cU_0(L) = \int_y^{y'}d\tl{y}\;\frac{b(\tl{y})}{n(\tl{y})}, \nonumber\\
 \cH_1(y,y') &\equiv g_{0,1}(y)-g_{L,1}(y')-\cG_1, \nonumber\\
 \cH_2(y,y') &\equiv g_{0,2}(y)-g_{0,1}(y)g_{L,1}(y')+g_{L,2}(y')-\cG_1\cH_1(y,y')-\cG_2, \nonumber\\
 \cH_3(y,y') &\equiv g_{0,3}(y)-g_{0,2}(y)g_{L,1}(y')+g_{0,1}(y)g_{L,2}(y')-g_{L,3}(y') \nonumber\\
 &\quad
 -\cG_1\cH_2(y,y')-\cG_2\cH_1(y,y')-\cG_3, \nonumber\\
 \cH_4(y,y') &\equiv g_{0,4}(y)-g_{0,3}(y)g_{L,1}(y')+g_{0,2}(y)g_{L,2}(y')
 -g_{0,1}(y)g_{L,3}(y')+g_{L,4}(y') \nonumber\\
 &\quad
 -\cG_1\cH_3(y,y')-\cG_2\cH_2(y,y')-\cG_3\cH_1(y,y')-\cG_4, \nonumber\\
 &\vdots
 \label{def:cHl}
\end{align}
From \eqref{tlG_B}, we also have
\begin{align}
 \der_y\der_y'\tl{G}_{\rm B<}(\rho^2,y,y') &\simeq \frac{e^{-\rho\Dlt(y,y')}}{2\asc^3n^{3/2}(y)n^{3/2}(y')}
 \sbk{\rho\cS_{-1}+\cS_0+\frac{\cS_{1}}{\rho}+\frac{\cS_{2}}{\rho^2}+\frac{\cS_{3}}{\rho^3}+\cdots}, 
 \label{der2tlG_B}
\end{align}
where
\begin{align}
 \cS_l(y,y') &\equiv -\frac{b(y)b(y')}{n(y)n(y')}\cH_{l+1}(y,y')
 +\brc{\frac{3}{2}\frac{n'(y)b(y')-b(y)n'(y')}{n(y)n(y')}-\frac{b(y')}{n(y')}\der_y+\frac{b(y)}{n(y)}\der_y'}\cH_{l}(y,y') \nonumber\\
 &\quad
 +\brc{\der_y\der_y'-\frac{3n'(y')}{2n(y')}\der_y-\frac{3n'(y)}{2n(y)}\der_y'+\frac{9n'(y)n'(y')}{4n(y)n(y')}}\cH_{l-1}(y,y'), 
\end{align}
with 
\begin{align}
 \cH_0(y,y') &\equiv 1, \;\;\;\;\;
 \cH_l(y,y') \equiv 0. \;\;\;\;\; \brkt{l<0}
\end{align}

\subsection{Spinor sector}
\subsubsection{Basis functions}
The basis functions~$C_u^{(\pm)}(y;\rho^2)$, $S_u^{(\pm)}(y;\rho^2)$ ($u=0,L$) are solutions of 
\begin{align}
 \sbk{\der_y^2+\brkt{\frac{n'}{n}-\frac{b'}{b}}\der_y-\frac{b^2}{n^2}\rho^2\pm\frac{n'}{n}bM-b^2M^2}f_u^{(\pm)}(y;\rho) &= 0. 
\end{align}
For large values of $\rho$, the solution~$f_u^{(\pm)}(y;\rho)$ of this equation can be expanded as
\begin{align}
 f_u^{(\pm)}(y;\rho) &= e^{\rho\cU_u(y)}\sbk{\frac{1}{2}+\frac{h_{u,1}^{(\pm)}(y)}{\rho}+\frac{h_{u,2}^{(\pm)}(y)}{\rho^2}
 +\frac{h_{u,3}^{(\pm)}(y)}{\rho^3}+\cdots}, 
\end{align}
where
\begin{align}
 h_{u,1}^{(\pm)\prime}(y) &= \frac{1}{4}\brc{\mp Mn'(y)+n(y)b(y)M^2}, \nonumber\\
 h_{u,l}^{(\pm)\prime}(y) &= -\frac{1}{2}\bigg[
 \frac{n(y)}{b(y)}h_{u, l-1}^{(\pm)\prime\prime}(y)+\brkt{\frac{n'(y)}{n(y)}-\frac{b'(y)}{b(y)}}\frac{n(y)}{b(y)}h_{u,l-1}^{(\pm)\prime}(y) \nonumber\\
 &\hspace{15mm}
 +\brc{\pm Mn'(y)-n(y)b(y)M^2}h_{u,l-1}^{(\pm)}(y)\bigg]. \;\;\;\;\; (l=2,3,4,\cdots)
\end{align}

Considering the conditions in \eqref{BC0:spinor} and \eqref{BCL:spinor}, 
the basis function~$C_u^{(\pm)}(y;\rho^2)$ is expressed as
\begin{align}
 C_u^{(\pm)}(y;\rho^2) &= f_u^{C(\pm)}(y;\rho)+f_u^{C(\pm)}(y;-\rho), 
\end{align}
where 
\begin{align}
 f_u^{C(\pm)}(y;\rho) &= e^{\rho\cU_u(y)}\sbk{\frac{1}{2}+\frac{h_{u,1}^{C(\pm)}(y)}{\rho}+\frac{h_{u,2}^{C(\pm)}(y)}{\rho^2}
 +\frac{h_{u,3}^{C(\pm)}(y)}{\rho^3}+\cdots}, 
\end{align}
and the functions~$h_{u,m}^{C(\pm)}(y)$ $(m=1,2,3,\cdots)$ are subject to the boundary conditions, 
\begin{align}
 h_{u,1}^{C(\pm)}(u) &= 0, \nonumber\\
 h_{u,2l}^{C(\pm)}(u) &= 0, \;\;\;\;\;
 h_{u,2l+1}^{C(\pm)}(u) = -\frac{n(u)}{b(u)}h_{u,2l}^{C(\pm)\prime}(u). \;\;\;\;\;
 \brkt{l=1,2,3,\cdots}
\end{align}

As for $S_u^{(\pm)}(y;\rho^2)$, it is expressed as
\begin{align}
 S_u^{(\pm)}(y;\rho^2) &= \frac{n(u)}{\rho b(u)}\sbk{f_u^{S(\pm)}(y;\rho)-f_u^{S(\pm)}(y;-\rho)}, 
\end{align}
where 
\begin{align}
 f_u^{S(\pm)}(y;\rho) &= e^{\rho\cU_u(y)}\sbk{\frac{1}{2}+\frac{h_{u,1}^{S(\pm)}(y)}{\rho}+\frac{h_{u,2}^{S(\pm)}(y)}{\rho^2}
 +\frac{h_{u,3}^{S(\pm)}(y)}{\rho^3}+\cdots}, 
\end{align}
and $h_{u,m}^{S(\pm)}(y)$ ($m=1,2,3,\cdots$) are subject to 
\begin{align}
 h_{u,2l-1}^{S(\pm)}(u) &= 0, \;\;\;\;\;
 h_{u,2l}^{S(\pm)}(u) = -\frac{n(u)}{b(u)}h_{u,2l-1}^{S(\pm)\prime}(u). \;\;\;\;\; \brkt{l=1,2,3,\cdots}
\end{align}

\subsubsection{5D propagator}
When $\bar{\rho}\abs{\cU_u(y)}\gg 1$, the functions in \eqref{def:H0L} are expanded as 
\ignore{
\begin{align}
 C_0^{(\pm)}(y;\rho^2) &= e^{\rho\cU_0(y)}\sbk{\frac{1}{2}+\frac{h_{0,1}^{C(\pm)}(y)}{\rho}+\frac{h_{0,2}^{C(\pm)}(y)}{\rho^2}
 +\frac{h_{0,3}^{C(\pm)}(y)}{\rho^3}+\cdots}, \nonumber\\
 S_0^{(\pm)}(y;\rho^2) &= \frac{n(0)}{b(0)}\frac{e^{\rho\cU_0(y)}}{\rho}
 \sbk{\frac{1}{2}+\frac{h_{0,1}^{S(\pm)}(y)}{\rho}+\frac{h_{0,2}^{S(\pm)}(y)}{\rho^2}+\frac{h_{0,3}^{S(\pm)}(y)}{\rho^3}+\cdots}, \nonumber\\
 C_L^{(\pm)}(y;\rho^2) &= e^{-\rho\cU_L(y)}\sbk{\frac{1}{2}-\frac{h_{L,1}^{C(\pm)}(y)}{\rho}+\frac{h_{L,2}^{C(\pm)}(y)}{\rho^2}
 -\frac{h_{L,3}^{C(\pm)}(y)}{\rho^3}+\cdots}, \nonumber\\
 S_L^{(\pm)}(y;\rho^2) &= -\frac{n(L)}{b(L)}\frac{e^{-\rho\cU_L(y)}}{\rho}
 \sbk{\frac{1}{2}-\frac{h_{L,1}^{S(\pm)}(y)}{\rho}+\frac{h_{L,2}^{S(\pm)}(y)}{\rho^2}-\frac{h_{L,3}^{S(\pm)}(y)}{\rho^3}+\cdots}. 
\end{align}
}
\begin{align}
 H_{0\chi}(y;\rho^2) &= \frac{e^{\rho\cU_0(y)}}{\rho}\sbk{\frac{1}{2}+\frac{h_{0,1}^{S(-)}(y)}{\rho}+\frac{h_{0,2}^{S(-)}(y)}{\rho^2}
 +\frac{h_{0,3}^{S(-)}(y)}{\rho^3}+\cdots}, \nonumber\\
 H_{L\chi}(y;\rho^2) &= -\frac{e^{-\rho\cU_L(y)}}{\rho}\sbk{\frac{1}{2}-\frac{h_{L,1}^{S(-)}(y)}{\rho}+\frac{h_{L,2}^{S(-)}(y)}{\rho^2}
 -\frac{h_{L,3}^{S(-)}(y)}{\rho^3}+\cdots}, \nonumber\\
 H_{0\lmd}(y;\rho^2) &= e^{\rho\cU_0(y)}\sbk{\frac{1}{2}+\frac{\tl{h}_{0,1}^{(+)}(y)}{\rho}+\frac{\tl{h}_{0,2}^{(+)}(y)}{\rho^2}
 +\frac{\tl{h}_{0,3}^{(+)}(y)}{\rho^3}+\cdots}, \nonumber\\
 H_{L\lmd}(y;\rho^2) &= e^{-\rho\cU_L(y)}\sbk{\frac{1}{2}-\frac{\tl{h}_{L,1}^{(+)}(y)}{\rho}+\frac{\tl{h}_{L,2}^{(+)}(y)}{\rho^2}
 -\frac{\tl{h}_{L,3}^{(+)}(y)}{\rho^3}+\cdots}, 
\end{align}
where
\begin{align}
 \tl{h}_{0,m}^{(+)}(y) &\equiv h_{0,m}^{C(+)}(y)-n(0)Mh_{0,m-1}^{S(+)}(y), \nonumber\\
 \tl{h}_{L,m}^{(+)}(y) &\equiv h_{L,m}^{C(+)}(y)-n(L)Mh_{L,m-1}^{S(+)}(y), \;\;\;\;\; \brkt{m=1,2,3,\cdots}
\end{align}
with $h_{0,0}^{S(+)}(y)=h_{L,0}^{S(+)}(y)=1/2$. 
Then, the 5D propagator is expanded as 
\begin{align}
 \bar{G}_{\chi\chi<}(\rho^2,y,y') &= -\frac{e^{-\rho\Dlt(y,y')}}{2\asc^3\rho}
 \sbk{1+\frac{2\cK_1^{\chi\chi}(y,y')}{\rho}+\frac{2\cK_2^{\chi\chi}(y,y')}{\rho^2}+\frac{2\cK_3^{\chi\chi}(y,y')}{\rho^3}+\cdots}, \nonumber\\
 \bar{G}_{\chi\lmd<}(\rho^2,y,y') &= \frac{e^{-\rho\Dlt(y,y')}}{2\asc^3}
 \sbk{1+\frac{2\cK_1^{\chi\lmd}(y,y')}{\rho}+\frac{2\cK_2^{\chi\lmd}(y,y')}{\rho^2}+\frac{2\cK_3^{\chi\lmd}(y,y')}{\rho^3}+\cdots}, \nonumber\\
 \bar{G}_{\lmd\chi<}(\rho^2,y,y') &= -\frac{e^{-\rho\Dlt(y,y')}}{2\asc^3}\sbk{1+\frac{2\cK_1^{\lmd\chi}(y,y')}{\rho}
 +\frac{2\cK_2^{\lmd\chi}(y,y')}{\rho^2}+\frac{2\cK_3^{\lmd\chi}(y,y')}{\rho^3}+\cdots}, \nonumber\\
 \bar{G}_{\lmd\lmd<}(\rho^2,y,y') &= -\frac{e^{-\rho\Dlt(y,y')}}{2\asc^3\rho}
 \sbk{1+\frac{2\cK_1^{\lmd\lmd}(y,y')}{\rho}+\frac{2\cK_2^{\lmd\lmd}(y,y')}{\rho^2}+\frac{2\cK_3^{\lmd\lmd}(y,y')}{\rho^3}+\cdots}, 
 \label{ap:barG}
\end{align}
where
\begin{align}
 \cK_1^{\chi\chi}(y,y') &\equiv h_{0,1}^{S(-)}(y)-h_{L,1}^{S(-)}(y')+h_{L,1}^{S(-)}(0), \nonumber\\
 \cK_m^{\chi\chi}(y,y') &\equiv h_{0,m}^{S(-)}(y)+(-1)^m\brc{h_{L,m}^{S(-)}(y')-h_{L,m}^{S(-)}(0)}
 +2\sum_{l=1}^{m-1}(-1)^l h_{0,m-l}^{S(-)}(y)h_{L,l}^{S(-)}(y') \nonumber\\
 &\quad
 -2\sum_{l=1}^{m-1}(-1)^{l}h_{L,l}^{S(-)}(0)\cK_{m-l}^{\chi\chi}(y,y'), \;\;\;\;\; \brkt{m\geq 2}
\end{align}
and
\begin{align}
 \cK_m^{\chi\lmd}(y,y') &\equiv \left.\cK_m^{\chi\chi}(y,y')\right|_{h_{L,l}^{S(-)}(y')\to \tl{h}_{L,l}^{(-)}(y')}, \nonumber\\
 \cK_m^{\lmd\chi}(y,y') &\equiv \left.\cK_m^{\chi\chi}(y,y')\right|_{h_{0,l}^{S(-)}(y)\to \tl{h}_{0,l}^{(-)}(y)}, \nonumber\\
 \cK_m^{\lmd\lmd}(y,y') &\equiv \left.\cK_m^{\chi\chi}(y,y')\right|_{h_{0,l}^{S(-)}(y)\to \tl{h}_{0,l}^{(-)}(y),\,h_{L,l}^{S(-)}(y')\to \tl{h}_{L,l}^{(-)}(y')}. 
 \;\;\;\;\; \brkt{m=1,2,3,\cdots}
\end{align}

\subsection{Expansion of $\Dlt(y,y')$ and $\cW(a,y)$} \label{cUcW_expansion}
When $y'=y+1/\Lct$, $\Dlt(y,y')$ defined in \eqref{def:cU} (or \eqref{def:cHl}) can be expanded as
\begin{align}
 \Dlt(y,y') &= \frac{b(y)}{n(y)\Lct}\sbk{1+\frac{X^{(1)}(y)}{\Lct}+\frac{X^{(2)}(y)}{\Lct^2}+\frac{X^{(3)}(y)}{\Lct^3}+\cdots}, 
\end{align}
where
\begin{align}
 X^{(l)}(y) &= \frac{n(y)}{(l+1)!b(y)}\der_y^l\brkt{\frac{b(y)}{n(y)}}. 
\end{align}
Specifically, we can express $X^{(l)}(y)$ as
\begin{align}
 X^{(1)} &= \frac{1}{2!}\brkt{\frac{b'}{b}-\frac{n'}{n}}, \nonumber\\
 X^{(2)} &= \frac{1}{3!}\brkt{\frac{b''}{b}-\frac{n''}{n}}-\frac{2n'}{3n}X^{(1)}, \nonumber\\
 X^{(3)} &= \frac{1}{4!}\brkt{\frac{b^{(3)}}{b}-\frac{n^{(3)}}{n}}-\frac{3n'}{4n}X^{(2)}-\frac{n''}{4n}X^{(1)}, \nonumber\\
 X^{(4)} &= \frac{1}{5!}\brkt{\frac{b^{(4)}}{b}-\frac{n^{(4)}}{n}}-\frac{4n'}{5n}X^{(3)}-\frac{3n''}{10n}X^{(2)}-\frac{n^{(3)}}{15n}X^{(1)}, \nonumber\\
 X^{(5)} &= \frac{1}{6!}\brkt{\frac{b^{(5)}}{b}-\frac{n^{(5)}}{n}}-\frac{5n'}{6n}X^{(4)}-\frac{n''}{3n}X^{(3)}-\frac{n^{(3)}}{12n}X^{(2)}-\frac{n^{(4)}}{72n}X^{(1)}, \nonumber\\
 &\vdots
\end{align}

Next, we expand $\cW(a,y)$ defined in \eqref{def:cW}. 
Since the incomplete gamma function~$\Gm(a,\dlt)$ is expanded as
\begin{align}
 \Gm(a,\dlt) &= \Gm(a)\sbk{1-\dlt^a e^{-\dlt}\brkt{\frac{1}{\Gm(a+1)}+\frac{\dlt}{\Gm(a+2)}+\frac{\dlt^2}{\Gm(a+3)}+\cdots}}, 
\end{align}
we obtain
\begin{align}
 \cW(a,y) &\equiv \frac{\Gm(a,\bar{\rho}\Dlt(y,y'))}{\Dlt^a(y,y')} \nonumber\\
 &= \Gm(a)\brkt{\frac{n}{b}\Lct}^a\sbk{1+\frac{Y^{(1)}}{\Lct}+\frac{Y^{(2)}}{2\Lct^2}+\frac{Y^{(3)}}{6\Lct^3}+\frac{Y^{(4)}}{24\Lct^4}
 +\frac{Y^{(5)}}{120\Lct^5}+\cdots} \nonumber\\
 &\quad
 -\frac{\Gm(a)\bar{\rho}^a}{\Gm(a+1)}\sbk{1+\cO\brkt{\frac{1}{\Lct}}}, 
 \label{expand:cW}
\end{align}
where $y'=y+1/\Lct$, and 
\begin{align}
 Y^{(1)} &\equiv -aX^{(1)}, \nonumber\\
 Y^{(2)} &\equiv -(a+1)X^{(1)}Y^{(1)}-2!X^{(2)}, \nonumber\\
 Y^{(3)} &\equiv -(a+2)X^{(1)}Y^{(2)}-2(2a+1)X^{(2)}Y^{(1)}-3!aX^{(3)}, \nonumber\\
 Y^{(4)} &\equiv -(a+3)X^{(1)}Y^{(3)}-3(2a+2)X^{(2)}Y^{(2)}-6(3a+1)X^{(3)}Y^{(1)}-4!aX^{(4)}, \nonumber\\
 &\vdots
\end{align}
\ignore{
\begin{align}
 Y^{(2)} &\equiv a\brc{(a+1)X^{(1)2}-2X^{(2)}}, \nonumber\\
 Y^{(3)} &\equiv -a\brc{(a+1)(a+2)X^{(1)3}-6(a+1)X^{(1)}X^{(2)}+6X^{(3)}}, \nonumber\\
 Y^{(4)} &\equiv a\big\{(a+1)(a+2)(a+3)X^{(1)4}-12(a+1)(a+2)X^{(1)2}X^{(2)} \nonumber\\
 &\hspace{10mm}
 +12(a+1)X^{(2)2}+24(a+1)X^{(1)}X^{(3)}-24X^{(4)}\big\}, \nonumber\\
 Y^{(5)} &\equiv -a\big\{(a+1)(a+2)(a+3)(a+4)X^{(1)5}-20(a+1)(a+2)(a+3)X^{(1)3}X^{(2)} \nonumber\\
 &\hspace{13mm}
 +60(a+1)(a+2)X^{(1)2}X^{(3)}+60(a+1)X^{(1)}\brc{(a+2)X^{(2)2}-2X^{(4)}} \nonumber\\
 &\hspace{13mm}
 -120(a+1)X^{(2)}X^{(3)}+120X^{(5)}\big\}, \nonumber\\
 &\vdots
\end{align}
}

\printbibliography

\end{document}